\documentclass[sigconf]{acmart}
\usepackage{amsmath}
\usepackage{multirow}

\AtBeginDocument{%
  \providecommand\BibTeX{{%
    \normalfont B\kern-0.5em{\scshape i\kern-0.25em b}\kern-0.8em\TeX}}}

\copyrightyear{2024}
\acmYear{2024}
\setcopyright{acmlicensed}\acmConference[UIST '24]{The 37th Annual ACM Symposium on User Interface Software and Technology}{October 13--16, 2024}{Pittsburgh, PA, USA}
\acmBooktitle{The 37th Annual ACM Symposium on User Interface Software and Technology (UIST '24), October 13--16, 2024, Pittsburgh, PA, USA}
\acmDOI{10.1145/3654777.3676352}
\acmISBN{979-8-4007-0628-8/24/10}

\begin{document}
\newcommand{\sys}{Patchview}
\newcommand{\persona}{Alex}
\newcommand{\added}[1]{#1}
\newcommand{\deleted}[1]{}

\newcommand{\john}[1]{\textbf{\textcolor{blue}{John: #1}}}
\newcommand{\inquote}[1]{``\textit{#1}''}
\title{\sys{}: LLM-Powered Worldbuilding with Generative Dust and Magnet Visualization}

\author{John Joon Young Chung}
\email{jchung@midjourney.com}
\affiliation{%
  \institution{Midjourney}
  \city{San Francisco}
  \state{CA}
  \country{USA}
}

\author{Max Kreminski}
\email{mkreminski@midjourney.com}
\affiliation{%
  \institution{Midjourney}
  \city{San Francisco}
  \state{CA}
  \country{USA}
}

\renewcommand{\shortauthors}{Chung and Kreminski.}

\begin{abstract}

Large language models (LLMs) can help writers build story worlds by generating world elements, such as factions, characters, and locations. However, making sense of many generated elements can be overwhelming. Moreover, if the user wants to precisely control aspects of generated elements that are difficult to specify verbally, prompting alone may be insufficient. We introduce \sys{}, a customizable LLM-powered system that visually aids worldbuilding by allowing users to interact with story concepts and elements through the physical metaphor of magnets and dust. Elements in \sys{} are visually dragged closer to concepts with high relevance, facilitating sensemaking. The user can also steer the generation with verbally elusive concepts by indicating the desired position of the element between concepts. When the user disagrees with the LLM's visualization and generation, they can correct those by repositioning the element. These corrections can be used to align the LLM's future behaviors to the user's perception. With a user study, we show that \sys{} supports the sensemaking of world elements and steering of element generation, facilitating exploration during the worldbuilding process. \sys{} provides insights on how customizable visual representation can help sensemake, steer, and align generative AI model behaviors with the user's intentions.
\end{abstract}



\begin{teaserfigure}
\includegraphics[width=\textwidth]{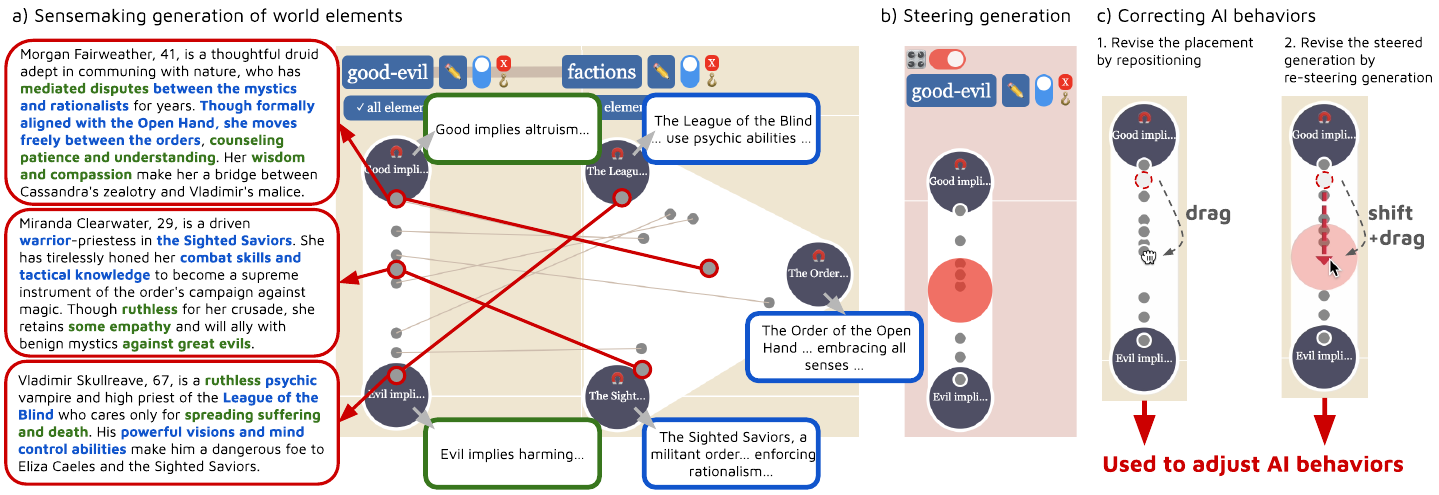}
  \caption{For generating story world elements with LLMs, \sys{} leverages a dust and magnet visual representation to help users a) sensemake, b) steer, and c) correct LLM generation. a) In dust and magnet visual representation, user-defined concepts serve as ``magnets’’ (larger dark circles), attracting ``dust particle’’ elements (smaller light circles) more strongly if an element is more relevant to the concept. Note that we only show partial excerpts of magnets due to the limited space. b) By placing a red marker between magnets, the user can steer the generation with a mix of different concepts. c) When LLM behaviors (steering, recognition) do not align with the user’s perception, the user can correct them simply by moving the element, either 1) revising the element position or 2) re-steering generation to rewrite the element. The corrected placement will be fed into the LLM pipeline as an example to improve future steering and recognition.}
  \Description{The figure demonstrates how LLMs can generate elements for a story world using a "dust and magnet" visual representation. The image is divided into three parts: a) Sensemaking generation of world elements: This section shows how LLM-generated story elements (represented as smaller light circles or "dust particles") are attracted to user-defined concepts (represented as larger dark circles or "magnets"). The relevance of an element to a concept determines how strongly it is attracted. Three character descriptions are provided as examples, with the two views. The first view has two concepts, "Good" and "Evil." The second view has three concepts of factions, "The League of the Blind... use psychic abilities...", "The Order of the Open Hand... embracing all senses...", and "The Sighted Saviors, a militant order... enforcing rationalism..." The first character is placed close to "Good" in the first view, and placed in the middle of three factions, with the description of "Morgan Fairweather, 41, is a thoughtful druid adept in communing with nature, who has mediated disputes between the mystics and rationalists for years. Though formally aligned with the Open Hand, she moves freely between the orders, counseling patience and understanding. Her wisdom and compassion make her a bridge between Cassandra's zealotry and Vladimir's malice." The second character is placed in the middle of "Good" and "Middle" and is close to "The Sighted Saviors...", with the description "Miranda Clearwater, 29, is a driven warrior-priestess in the Sighted Saviors. She has tirelessly honed her combat skills and tactical knowledge to become a supreme instrument of the order's campaign against magic. Though ruthless for her crusade, she retains some empathy and will ally with benign mystics against great evils." The third character is close to "Evil" and "The League of the Blind...", with the description of "Vladimir Skullreave, 67, is a ruthless psychic vampire and high priest of the League of the Blind who cares only for spreading suffering and death. His powerful visions and mind control abilities make him a dangerous foe to Eliza Caeles and the Sighted Saviors." b) Steering generation: In this part, the visual spaces between the concept "magnets" are used to steer the LLM generation. The user can specify weights for the considered concepts (indicated by a red circle) to influence the generation process. In the example, the red circle is placed in the middle between "Good implies..." and "Evil implies..." c) Correcting AI behaviors: This section illustrates how users can correct misalignments between LLM behaviors and their own perceptions. The user can either 1) revise the element's placement directly or 2) re-steer the generation by re-positioning the element. The corrections are then fed back into the LLM pipeline as examples for future steering and recognition tasks, which is noted with "Used to adjust AI behaviors."}
  \label{fig:teaser}
\end{teaserfigure}


\maketitle

\section{Introduction}
Rapid progress in the development of generative large language models (LLMs)~\cite{brown2020language, chowdhery2022palm} has recently led to the introduction of numerous LLM-based tools for storywriting~\cite{chung2022talebrush, yuan2022wordcraft, sudowrite_2023, lee2024dsiiwa}. While many of these tools aim to generate text for direct inclusion in a finished story, opportunities also lie in using LLMs to support other aspects of the writing process, such as worldbuilding. Worldbuilding---the act of constructing a coherent fictional world~\cite{hergenrader2018collaborative}---establishes a setting from which a variety of stories could arise. It requires writers to envision myriad aspects of a world, from abstract values (e.g., religion, ideology) to more specific elements, such as factions, characters, places, or props. As worldbuilding involves creating many different world elements, writers often put a lot of time and effort into it. Generative LLMs could be used to support this process, for instance by producing additional world elements that fit into the established setting or even inspire writers to expand their conception of the world they are creating.

However, when generating many world elements with LLMs, understanding their overall landscape can be challenging. That is, to unfold a story where different elements interact with each other, the writers would need to have a holistic view regarding what kind of attributes and values those elements hold. As LLMs can quickly add many elements to the world, it can be challenging for a writer to understand the rapidly growing world. Moreover, once the writer has understood existing world elements, they might want to generate a specific type of world elements. One way to guide LLMs for such a purpose would be to write natural language prompts. However, if the writer wants to express verbally elusive or ambiguous concepts, writing natural language prompts can either be cumbersome~\cite{subramonyam2024bridging} or have limited expressivity~\cite{chung2023promptpaint}. 

To support sensemaking and steering of world element generation, 
\added{we propose \textit{generative dust and magnet} (GD\&M) visualization, which adapts dust and magnet visual representation~\cite{yi2005dust, chen2018anchorviz} to the use of generative models.}\deleted{we propose to adopt a dust and magnet visual representation.} 
\added{GD\&M}\deleted{Dust and magnet representations} visualizes elements as ``particles of iron dust'' which are attracted to different concepts, or  ``magnets'', based on their relevance to each concept (i.e., placed more closely if more relevant) (Figure~\ref{fig:teaser}a). This approach supports flexible visualization of semantic association between concepts and elements with an arbitrary number of concepts, by leveraging intermediate spaces between extreme ``anchoring'' concepts. Moreover, spaces between concepts can be used for guiding generation, even allowing expression of ambiguity between concepts (Figure~\ref{fig:teaser}b). When the user disagrees with steered generation and recognition results, the user can straightforwardly correct them by simply moving dust particles to other positions (Figure~\ref{fig:teaser}c). With repositioning, the user can indicate the generated element's correct placement (Figure~\ref{fig:teaser}c1) or command AI to revise the element to fit in the new position (Figure~\ref{fig:teaser}c2).
These corrections can feed back into the AI as examples of the user's perspective for future steering and recognition. 

We instantiated these interactions in \sys{}, an LLM-powered story worldbuilding tool. 
Via a user study with eight hobbyists and one professional in worldbuilding, we show that \sys{} allows users to quickly understand the landscape of elements within the story world. Moreover, we find that visual steering of LLM generation could function as an intuitive alternative to natural language prompting, allowing users to express nuances that are difficult to articulate with natural language. While participants found the interaction of correcting AI results on the visual space straightforward, user-provided corrections did not have a significant impact on aligning AI behaviors to user intent. However, participants found the tool overall helpful for worldbuilding, flexibly creating worlds that they found to be of interest.
We conclude with discussions on visual representations for interacting with generative AI; using worlds from \sys{} for story writing; technical alternatives for prompt engineering and closed models; and limitations. 
\deleted{In summary, this work introduces novel visual interaction approaches to make sense of, guide, and intervene in the use of generative AI models.}

\added{In summary, this work has three main contributions:}
\begin{enumerate}
\item \added{Generative Dust and Magnet (GD\&M) visualization, a novel visual interaction approach to make sense of, guide, and intervene in the use of generative AI models.}
\item \added{\sys{}, an LLM-powered tool that supports story world element creation with GD\&M.}
\item \added{An evaluation that shows how \sys{} supports sensemaking and steering of LLM outputs while revealing limitations in aligning LLM behaviors to the user's perspective.}
\end{enumerate}

\section{Related Work}

\subsection{Worldbuilding}

Worldbuilding is a process of architecting fictional worlds that can be cornerstones of narrative fiction~\cite{hergenrader2018collaborative}. It considers various aspects, such as places, characters, or even cultures, and well-constructed worlds add believability to the stemming narrative stories. A well-built story world also entertains readers, as readers build out the conception of a coherent world out of various stemming stories~\cite{maj2015transmedial, fast2017transmedia}. With a story world, readers can also participate in active consumer experiences, such as creating fan fiction and even transforming the canon world into alternative worlds~\cite{samutina2016fan, fast2017transmedia}. While worldbuilding can be a complex process with many aspects to consider, there have been practical frameworks and structures that practitioners use. Practitioners would likely first focus on the frameworks of the world, which can include scope (geography of the world), sequence (temporal history of the world), and perspective (from whom the world is explained)~\cite{hergenrader2018collaborative}. Under such frameworks, practitioners would create structures of the world. Governance (e.g., government presence, rule of law), economics (e.g., economic strength, wealth distribution), social relations (e.g., class, race, and ethnic relations), cultural influences (e.g., religious influences, cultural influences), and character alignments (e.g., good-evil, lawful-chaotic)~\cite{hergenrader2018collaborative} are some examples of world structures. Then, around frameworks and structures, practitioners would create catalogs of fictional worlds, or elements of the world, such as characters, places, props, and events~\cite{hergenrader2018collaborative}. Worldbuilding can be done either solely (e.g., Tolkien’s world of Lord of Rings) or collectively (e.g., Marvel Universe), and commercial projects often tend to be collaborative as doing creative work can be overloading to an individual. WorldSmith is one of the few AI-powered tools to support worldbuilding but focuses on creating visual aspects of the world~\cite{dang2023worldsmith}.
In this work, \added{which focuses on supporting world element creation,} we introduce an AI-powered worldbuilding tool that co-constructs the story world with users by generating new world elements based on what the user has. Specifically, we facilitate the use of AI models by incorporating visual means for users to sensemake and control world element generation. 

\subsection{AI-Powered Story Writing}

With advancing AI technologies, researchers and practitioners have developed many tools to support story writing. For example, TaleStream supports story ideation by showing potentially inspiring story tropes~\cite{chou2023talestream}. Loose Ends is a rule-based mixed-initiative AI system that allows users to explore plot threads with some constraints~\cite{kreminski2022loose}. Portrayal leveraged NLP and visualizations to help writers analyze characters in their stories~\cite{hoque2023portrayal}. LLM's advanced generative capabilities introduced tools that suggest texts that users can incorporate into their writing~\cite{clark2028creative, lee2022coauthor, calderwood2020hownovelists}. Researchers investigated diverse interactions for such tools, from allowing distinct suggestion operations~\cite{yuan2022wordcraft} to incorporating multimodality~\cite{gong2023interactive}, hierarchical generation~\cite{mirowski2023cowriting}, and sketching inputs~\cite{chung2022talebrush}. With these rapidly advancing capabilities, researchers also studied story writer’s expectations for these technologies, such as what they would take as a benefit and what they want~\cite{biermann2022from, gero2023social, kim2024authors}. 
Lee et al.~\cite{lee2024dsiiwa} reflected on the design space of writing tools through a literature survey. 
LLMs also enabled story applications where the story is generated with minimal writer interventions, directly facing the audience~\cite{aidungeon2_2019, park2023generative}. While many LLM-powered story writing tools focused on supporting prose text writing, some focused on other types of support. For example, CALYPSO leverages LLMs to provide support to dungeon masters when playing Dungeons \& Dragons~\cite{zhu2023calypso}. In a similar vein, we design \sys{} to provide LLM-powered support in worldbuilding, which is other than writing story texts themselves. 

\subsection{Visually Interacting with Generative AI}
\label{sec:rw_visual}
While natural language-based interfaces (e.g., prompts, chat) have been widely used for generative AI models, many previous systems used visual interactions to complement natural language interactions. Some tools leverage node-based input interactions to control generation, such as chaining subtasks~\cite{wu2022AIChains, arawjo2023chainforge, kim2023cells, angert2023spellburst}. Among them, ChainForge~\cite{arawjo2023chainforge} and Cells-Generators-Lenses framework~\cite{kim2023cells} also allowed evaluation of generated results with visualization nodes. While these tools allowed flexible control, steering and evaluation happened in separated interfaces, leading to visual complexity. As another type, Scenescape~\cite{suh2023sensecape} and Graphologue~\cite{jiang2023graphologue} leveraged graph and tree visualization to help understand complex information. While the user can steer further generations by clicking on the node which the user is willing to learn more details about, these focus more on presenting information than allowing flexible steering. Some tools allow steering or evaluation of multiple generation results on dimensional spaces of attributes, represented in either \added{sliders~\cite{louie2020novice},} mixed color spaces~\cite{chung2023promptpaint}, temporal line drawing~\cite{chung2022talebrush}, or scatter plots organized in grids~\cite{suh2023structured}. Among them, TaleBrush~\cite{chung2022talebrush} and Luminate~\cite{suh2023structured} tied steering and evaluation interactions on a single visual representation, minimizing clutters. TaleBrush considers a continuous dimensional scale but on a fixed attribute. On the other hand, Luminate allows arbitrary dimensional attributes but only with categorical/ordinal attributes. Moreover, all aforementioned tools do not allow users to correct AI behaviors when AI’s steered generation and recognition results do not align with the user's thoughts. \sys{} extends previous work by allowing generation steering, evaluation, and user corrections on an integrated single visual representation with the flexibility of allowing continuous scales of any arbitrary concepts of interest. 
\section{Generative Dust and Magnet}
\label{sec:design}
\sys{}'s central design metaphor---\emph{generative dust and magnet} (GD\&M)---leverages a dust and magnet (D\&M) visual representation~\cite{yi2005dust} to facilitate interaction with generative AI models. In this section, we first describe settings where GD\&M can be helpful (Section~\ref{sec:design_need}). Then, we describe the original D\&M visualization and how we translate its components for use with generative models (Section~\ref{sec:design_dustnmagnet_translation}). Finally, we describe specific GD\&M interactions that close gaps in the interactive alignment of AI models: evaluation support, specification alignment, and process alignment~\cite{terry2023ai} (Section~\ref{sec:interaction_alignment}). 

\subsection{Need for Generative Dust and Magnet}
\label{sec:design_need}

Interaction with generative AI might benefit from a wide range of different interaction approaches in different settings. In general, we expect GD\&M interaction to be most effective when \textbf{the user must generate many distinct units of output (e.g., storyworld elements) that vary along diverse and expressive conceptual dimensions}. Breaking this ideal setting down further, we arrive at a set of three conditions that typify good application domains for GD\&M interaction.

First, the user must make use of generative models to gather a collection of many generated outputs. This imposes a need for \textbf{sensemaking} (\textbf{N1}), as understanding how outputs distribute along the user's conceptual dimensions of interest is difficult due to the large scale of generation.

Second, the user must have desires to create artifacts within their unique characteristics and values, which often occurs in artistic creation~\cite{chung2022artist}.
This imposes a need for \textbf{configurability} (\textbf{N2}), where behaviors of AI functions (e.g., generation and evaluation of generated results) consider the user’s unique styles and interests. 

Third, the user must need to express nuanced specifications that align generation with the user’s specific intentions and facilitate exploration of subtly different options. This imposes a need for \textbf{expressivity} (\textbf{N3}) where the user can guide generation even with subtle intentions. 

GD\&M interaction would be ideal for user tasks with the above characteristics. Worldbuilding meets all of these conditions: the writer must create many world elements to fill out a unique and idiosyncratic world, and created elements can have nuanced differences between each other~\cite{hergenrader2018collaborative}. In the following sections, we describe how GD\&M can fulfill the aforementioned needs. 

\subsection{From D\&M Visualization to GD\&M}
\label{sec:design_dustnmagnet_translation}
\citeauthor{yi2005dust}'s original dust and magnet visualization represents individual data elements as ``dust particles'' while representing each variable for which data elements can possess different values as a ``magnet''. Both dust particles and magnets are rendered as glyphs on a 2D plane; a data element with a particularly high value for a certain variable will be placed closer to the magnet representing that variable. This approach can facilitate the accessibility of understanding many multivariate data instances~\cite{yi2005dust} while allowing users to identify notable patterns within a dataset~\cite{chen2018anchorviz}. 

\begin{figure}
    \includegraphics[width=0.478\textwidth]{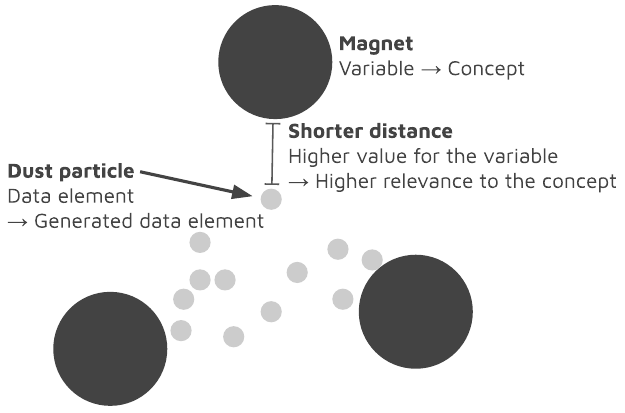}
    \caption{Compared to dust and magnet visualization, generative dust and magnet replaces data elements (dust particles) and variables (magnets) with generated data elements and concepts, respectively. In generative dust and magnet, the distance between a magnet and a dust particle indicates the intensity of relevance between them.}
    \Description{Diagram showing the relationship between magnets and dust particles and how they are interpreted for dust and magnet visualization and generative dust and magnet. For dust and magnet visualization, a magnet, which is a dark large circle, stands for a variable. A dust particle stands for a data element, and the distance between the magnet and the dust particle stands for the relevance between the variable and the data element. The shorter the distance, the data element has a higher value for the variable. When dust and magnet visualization translates to generative dust and magnet, variable and data element change to concept and generated data element, respectively. The distance between the magnet and dust particle stands for the relevance between the generated data element to the concept, with the shorter distance indicating higher relevance.}
    \label{fig:dustnmagnet_translation}
\end{figure}

We extend D\&M visualization to an interface for generative models (Figure~\ref{fig:dustnmagnet_translation}). Generative D\&M replaces multivariate data elements with generated data elements in the output modality of a generative model (e.g., passages of text for LLMs, and images for text-to-image models). Accordingly, variables in the original D\&M visualization translate to concepts that characterize the generated outputs \added{(e.g., ``positive sentiment'' for texts, ``pastel colors'' for images)}. Under this translation, a generated element that is more strongly relevant to a specific concept is drawn closer to the magnet for the corresponding concept.

\subsection{Specific GD\&M Interactions}
\label{sec:interaction_alignment}

Several specific GD\&M interactions are designed to meet user needs discussed in Section~\ref{sec:design_need} \added{(Figure~\ref{fig:dustnmagnet_interactions0})}. We organize these interactions in terms of how they support interactive alignment of AI models~\cite{terry2023ai}. Extending challenges of the gulf of evaluation and execution~\cite{norman2002design},
Terry et al.~\cite{terry2023ai} emphasized three facets of interactive alignment of AI models: 1) evaluation support (\textbf{I1}), or users making sense of AI outputs; 2) specification alignment (\textbf{I2}), or users efficiently and reliably communicating their objectives to AI; and 3) process alignment (\textbf{I3}), or users verifying or controlling AI’s execution process.

\subsubsection{I1: Evaluation Support - User Configurable Dust and Magnet Visualization (N1, N2)}

\begin{figure}
    \includegraphics[width=0.478\textwidth]{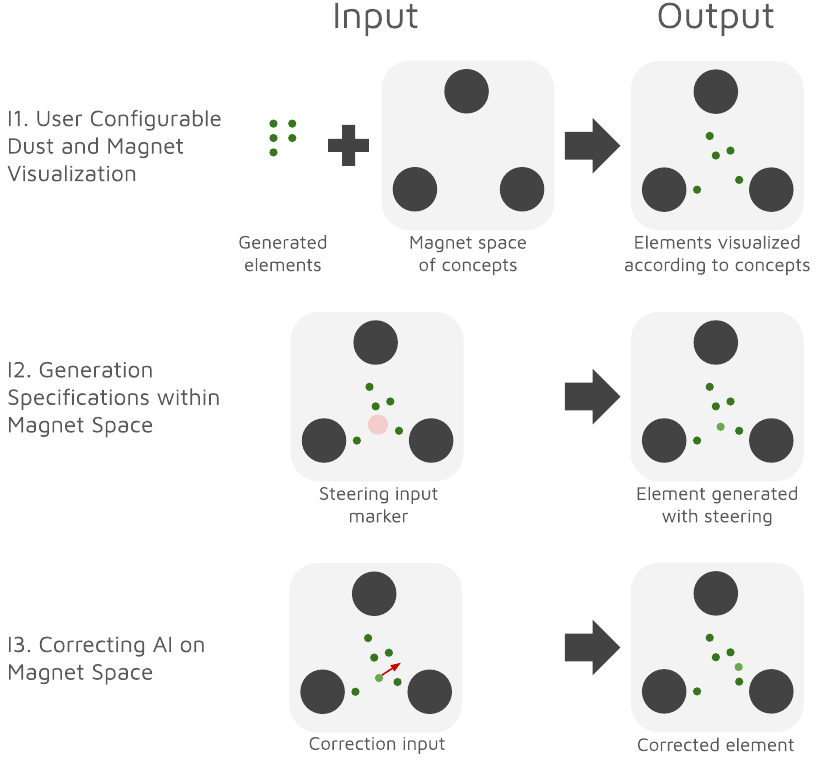}
    \caption{\added{Input-Output schemes for GD\&M Interactions}}
    \Description{Diagram illustrating the input and output flow for three interactions of GD\&M. The first subfigure shows the interaction for "User Configurable Dust and Magnet Visualization." In this interaction, generated elements and magnet space of concepts are inputs, which are represented as a set of small green circles and larger grey circles, respectively. Concepts are arranged in a triangular shape. The output resulting from these inputs is elements visualized according to concepts, where small green circles (elements) are positioned in between larger grey circles (concepts). The second subfigure shows the interaction for "Generation Specifications within Magnet Space," where the input is a steering input marker, which is denoted as a pink marker put in between grey circles for concepts. The output is an element generated with steering, which is denoted with a lighter green circle. It is positioned in a similar place as the pink marker in the input diagram. Note that the pink marker is not shown in the output diagram. The third subfigure shows the interaction for "Correcting AI on Magnet Space," whose input has a red arrow from the lighter green circle. The output diagram shows the lighter green circle moved to the position indicated in the input diagram. Note that the red arrow is not shown in the output diagram.}
    \label{fig:dustnmagnet_interactions0}
\end{figure}

\begin{figure}
    \includegraphics[width=0.478\textwidth]{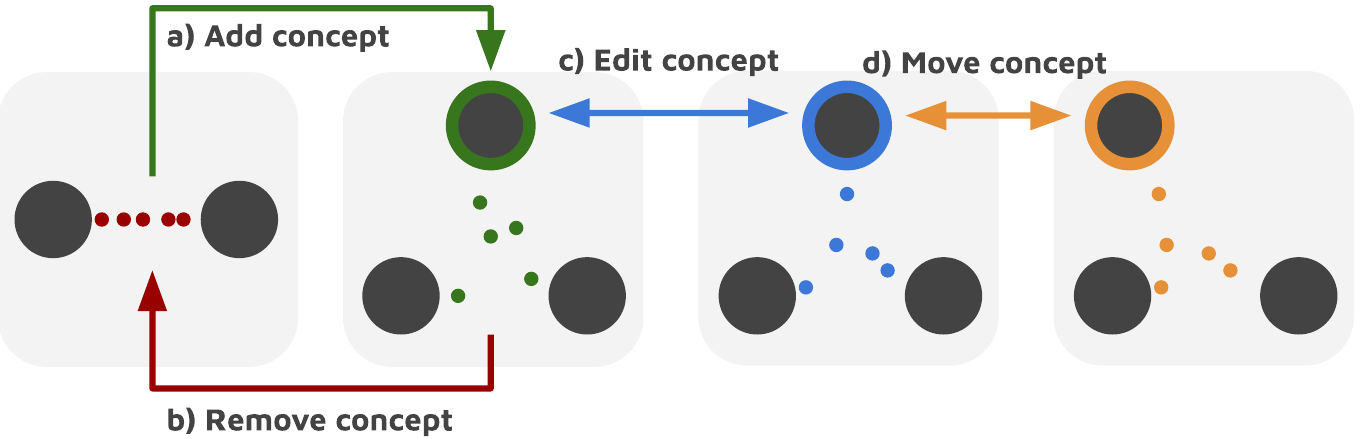}
    \caption{Configuration interactions for evaluation support in generative dust and magnet. As the user adds, removes, edits, and moves concepts according to how they want to organize elements and concepts, the positions of data elements get updated. }
    \Description{Diagram showing four configuration interactions in the generative dust and magnet visualization. Part a) shows adding a new concept, represented by a green circle, which attracts nearby gray dots representing data elements. Part b) shows removing a concept, with the corresponding green circle disappearing and the surrounding gray dots redistributing. Part c) depicts editing a concept, with the circle color changing from blue to green. The dot distribution changes accordingly. Part d) illustrates moving a concept, with the blue circle shifting position to the orange circle and the gray dots following its movement to maintain proximity to the concept.}
    \label{fig:dustnmagnet_interactions}
\end{figure}

To support users' sensemaking of many generated elements according to their concepts of interest, \added{the user can add generated elements to the magnet space configured with concepts of the user's interest (Figure~\ref{fig:dustnmagnet_interactions0}-I1). Then,}
an AI model measures the relevance of each element to different concepts and visualizes this information as the relative distance to those concepts \added{(e.g., good characters being closer to the concept of ``good'' than to ``evil'')}. With this support, the user can quickly get an overview of how generated elements are different from each other. GD\&M further provides users with flexible configurability, as the user can add, remove, or even edit concepts. With this configurability (Figure~\ref{fig:dustnmagnet_interactions}a, b, and c), users can easily reassess the set of generated outputs in terms of the concepts that are most relevant to their current focus. Moreover, the user can adjust the layout of concepts (Figure~\ref{fig:dustnmagnet_interactions}d), aligning their visual presentation with how they want to think about these concepts and elements. 

\subsubsection{I2: Specification Alignment - Generation Specifications within Magnet Space (N2, N3)}

GD\&M interaction also allows users to guide the generation of new elements by indicating the ideal placement of these elements within the visual-semantic magnet space defined by a set of user-configured concepts. That is, the user can place a marker on the magnet space to request the generation of elements that would be placed near the specified marker (as in Figure~\ref{fig:teaser}b \added{and~\ref{fig:dustnmagnet_interactions0}-I2}). \added{For instance, if the user wants to generate a good character, in the magnet space between ``good'' and ``evil,'' they can place the marker closer to ``good.''} One benefit of this visual magnet space is that the user can express vague or ambiguous specifications in this continuous space between concepts \added{(e.g., generating an array of characters that vary subtly along the ``good''-``evil'' spectrum)}.

\subsubsection{I3: Process Alignment - Correcting AI on Magnet Space (N2)}

AI behaviors may not always align with user intent: for instance, the user might not agree with how the AI interprets concepts during generation and placement of generated elements. In such cases, the user can freely re-specify concepts to more accurately convey how they think about each concept (\added{e.g., adding more details about what ``good'' means in a specific story world,} Figure~\ref{fig:dustnmagnet_interactions}c). They can also leverage the magnet space itself to correct AI behavior, by simply moving a misplaced generated element to wherever the user thinks it should be in the magnet space (Figure~\ref{fig:teaser}c \added{and~\ref{fig:dustnmagnet_interactions0}-I3}). Repositioning an element can convey two intentions: either 1) that the element's ``correct'' placement is in a new position (\added{e.g., indicating that the character should sit in the middle of ``good'' and ``evil'' as in }Figure~\ref{fig:teaser}c1) or 2) that the element should be revised to better fit the indicated position (\added{e.g., request AI to rewrite the character description to sit in the middle of ``good'' and ``evil'', as in }Figure~\ref{fig:teaser}c2). These corrections can then be used as examples to better align future generation and placement with user perception of concepts.
\section{\sys{}: Interface and Technical Details}

\begin{figure*}
    \includegraphics[width=\textwidth]{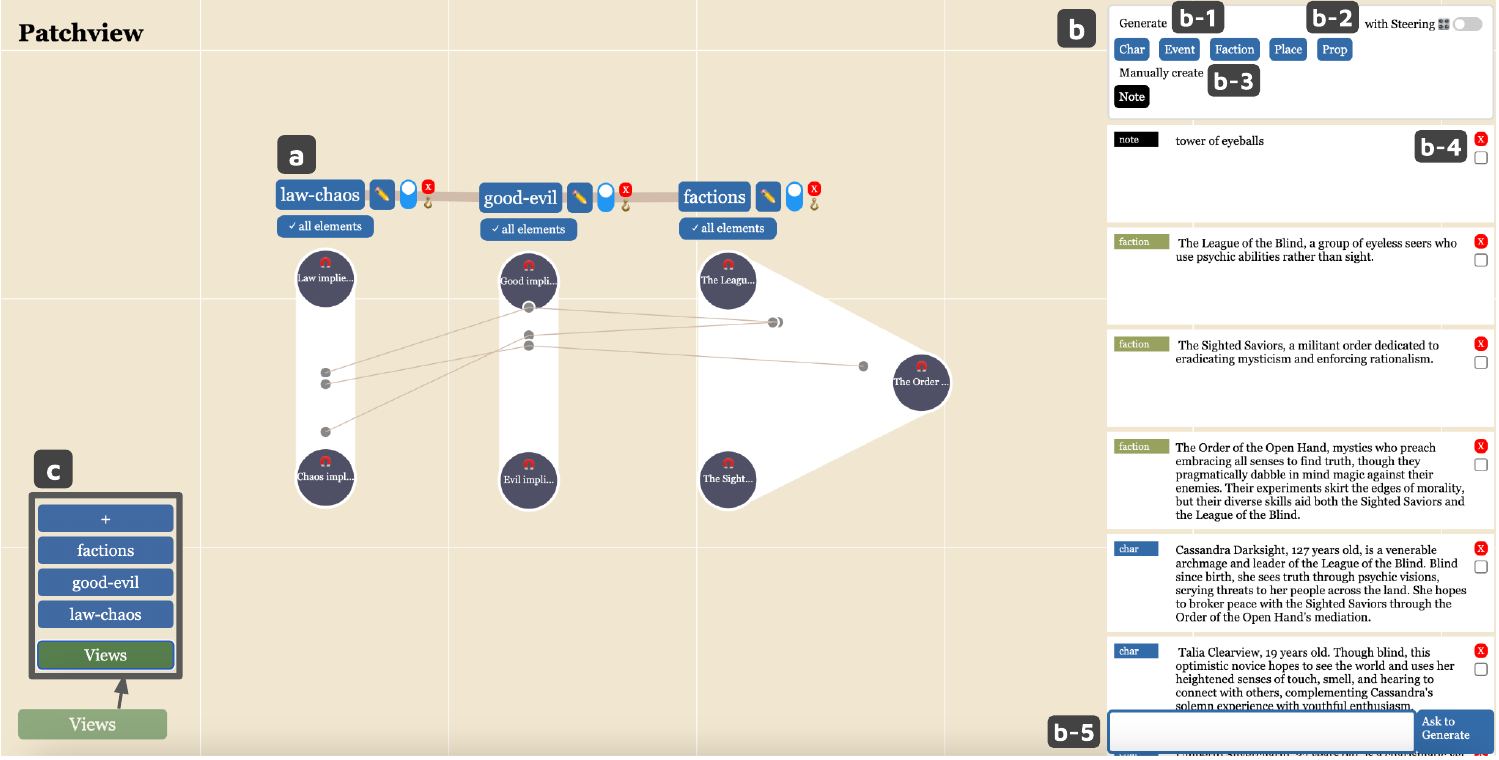}
    \caption{\sys{} interface. a) View module visualizes world elements in relation to concepts of the user’s interest. Specific interactions are shown in Figure~\ref{fig:teaser}. b) List module lists world elements as notes (b-4). This module allows users to generate elements by clicking buttons for different element types (b-1) or by prompting an LLM with specific natural-language instructions (b-5). The user can steer generation with a view interface (as in Figure~\ref{fig:teaser}b) by entering the steering mode with a toggle switch (b-2). They can also manually create notes (b-3). c) The user can see a list of existing views by clicking the Views button and create a new view with the + button.}
    \Description{The image shows the Patchview interface, which consists of two main modules: a View module (labeled 'a') and a List module (labeled 'b' and 'c'). The View module (a) visualizes world elements according to concepts of interest to the user. It displays views labeled "law-chaos", "good-evil", and "factions", which are connected. Under each view, there are concepts and elements, which are denoted with large and small circles, respectively. The List module (b) has elements as textual notes (b-4). At the top of the interface, there are buttons for generating different types of elements including characters, events, factions, places, and props (b-1). There is a text area at the bottom of the module, where the user can prompt language models to generate elements (b-5). At the top right of the module, there is a toggle switch (b-2), which lets users enter a steering mode to guide generations through a visual interface (shown in a separate figure). Users can manually create notes as well with the black "note" button (b-3). Existing views can be accessed by clicking the "Views" button (c), and new views can be created with the "+" button.}
    \label{fig:patchview_interface}
\end{figure*}

With GD\&M, we built \sys{}, an \deleted{AI}\added{LLM}-powered \deleted{story worldbuilding} tool \added{for world element creation} (Figure~\ref{fig:patchview_interface}). \deleted{Powered by LLMs}\added{Specifically}, \sys{} supports sensemaking and steering of world element generation. To demonstrate the effectiveness of GD\&M for sensemaking and steering, \sys{} focuses on creating initial ``seeds'' of story world elements in two to three sentences. Afterward, users can develop details of these seed elements either by themselves or with the help of AI; the final rendering of seed elements into a more complete form is left to future work.  

\sys{}'s user interface consists of the list module, which shows existing world elements as a list of notes (Figure~\ref{fig:patchview_interface}b), and the view module, which organizes world elements via GD\&M (Figure~\ref{fig:patchview_interface}a). Note that \sys{} leverages AI to generate specific world elements (e.g., characters, places) rather than generating frameworks or structures of the world (e.g., ideology, values). The user can manually specify frameworks and structures as open-ended text in notes.

We explain the envisioned usage pattern with a hypothetical user, \persona{}. \persona{} is a game scenario writer who is trying to design a story world for the new game her team is developing. To get help with the process, \persona{} decides to use \sys{}. 

\subsection{List Module}

\begin{figure}
    \includegraphics[width=0.4\textwidth]{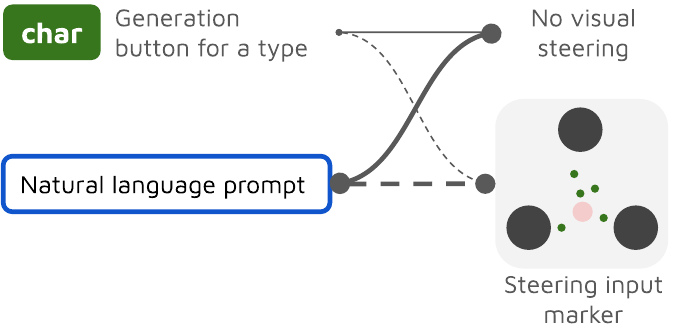}
    \caption{\added{Possible inputs to generate elements.}}
    \Description{The left side of figure has two items, ``Generation button for a type'' and ``Natural language prompt.'' The right side has two items, also, ``No visual steering'' and ``steering input marker.'' The left top and right top are connected with a thin solid line. The left top and right bottom are connected with a thin dashed line. The left bottom and right top are connected with a thick solid line. The left bottom and right bottom are connected with a thick dashed line.}
    \label{fig:patchview_gen-path}
\end{figure}

As \persona{} loads \sys{}, she first sees the list module on the right. With this module, \persona{} can generate and create an initial set of world elements as textual notes. To set an initial high-level concept for the world, \persona{} decides to manually create a note by clicking the \texttt{note} button (Figure~\ref{fig:patchview_interface}b-3) and modifying the text to “tower of eyeballs.” Next, extrapolating from this high-level idea, \persona{} decides to generate factions in the story world. Generating world elements with AI is straightforward, as \persona{} can simply click on the button that corresponds to the type of the element that \persona{} wants to introduce (Figure~\ref{fig:patchview_interface}b-1, \added{thin solid line in Figure~\ref{fig:patchview_gen-path}}). When generating the new element, by default, \sys{} will take all existing elements into context to ensure that the generated element is relevant to the current story world. If \persona{} wants the generation to consider only a subset of existing elements, she can select only those notes as context for generation. In case \persona{} wants to generate a more specific world element, \persona{} can also directly prompt the AI with natural language (Figure~\ref{fig:patchview_interface}b-5, \added{thick solid line in Figure~\ref{fig:patchview_gen-path}}). With this generation function, \persona{} first generates a few factions and then a handful of characters. As the number of world elements increases, \persona{} \added{can organize them in the list by reordering them with dragging. However, at a certain point, she }feels that the list is getting longer and becoming hard to understand. 

\subsection{View Module}

\begin{figure}
    \includegraphics[width=0.478\textwidth]{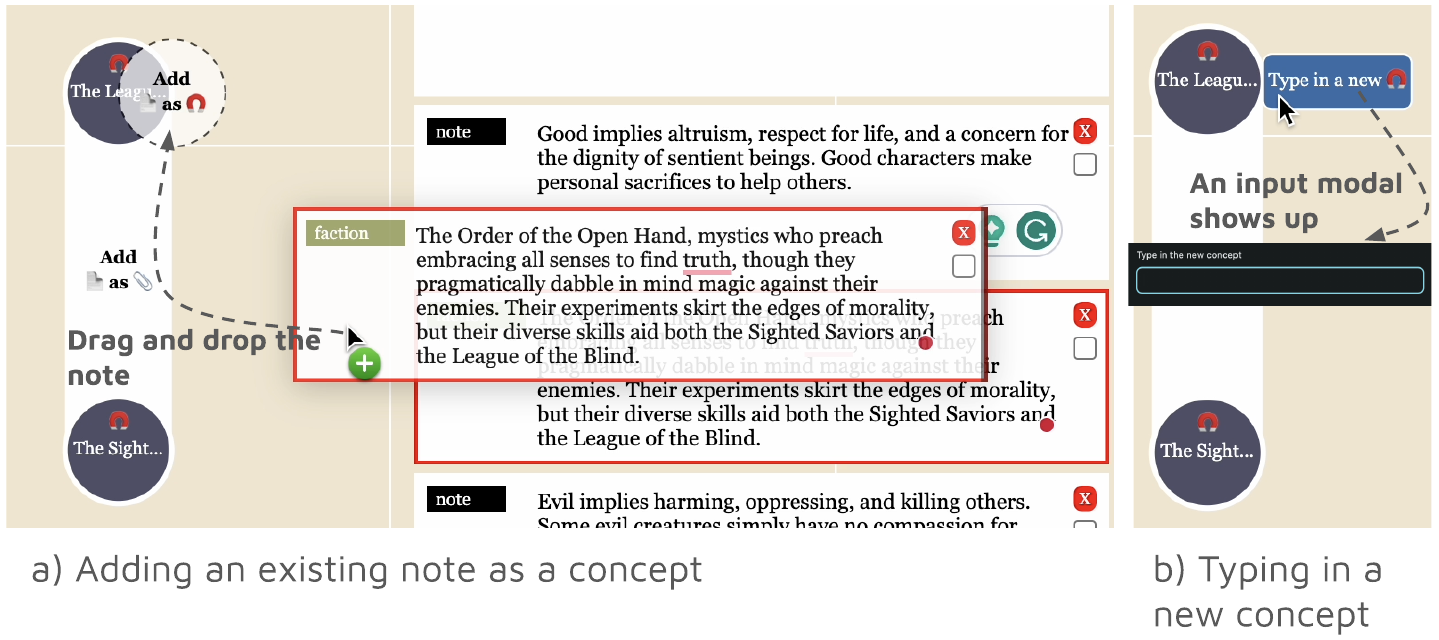}
    \caption{Interactions to add a new concept to the view.}
    \Description{The image shows interactions to add a new concept to a view in a user interface. On the left, there is a screenshot of the tool with the title "a) Adding an existing note as a concept". The user is dragging a note from the list of notes to the placeholder on the view saying "Add note as magnet." On the right, there is a figure titled "b) Typing in a new concept." In the figure, the user clicks "Type in a new magnet" button and the text input modal shows up.}
    \label{fig:patchview_concept_add}
\end{figure}

\begin{figure}
    \includegraphics[width=0.35\textwidth]{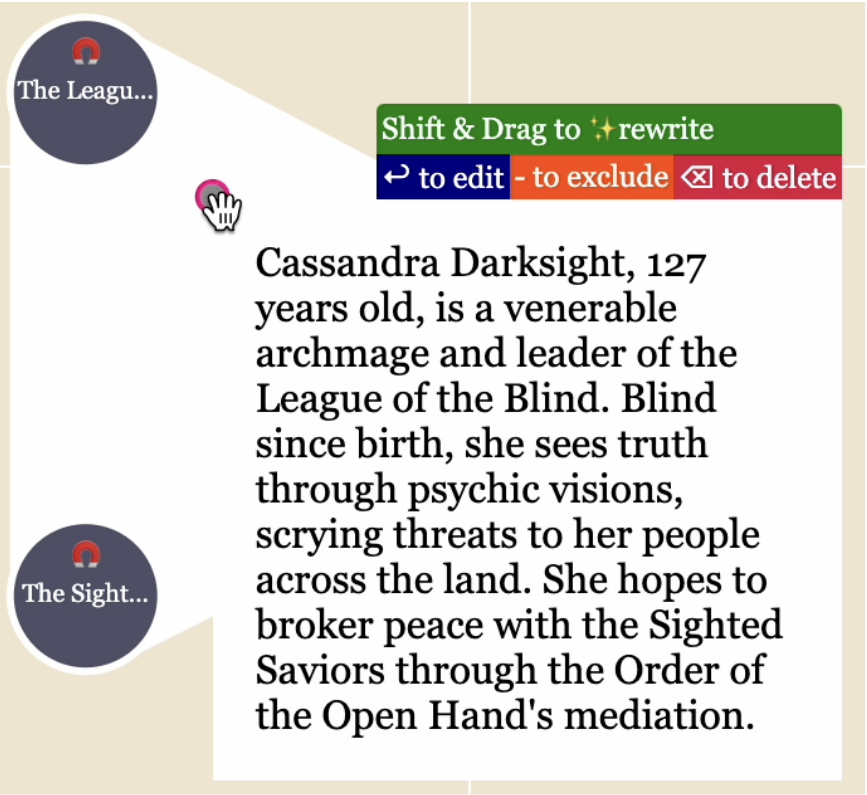}
    \caption{The user can read each concept and element by hovering their mouse over them. While hovering the mouse, they can 1) edit them by hitting the enter key, 2) exclude them from the view by hitting the - key, and 3) delete them by hitting backspace. For elements, the user can re-steer the generation by dragging the element to a new position while holding the shift key; the LLM will then attempt to rewrite the element's text to better match the target position.}
    \Description{The figure shows the user hovering their mouse over an element in the view, and the interface shows the textual content of the element. The element is labeled with the following possible commands: 1) shift and drag to rewrite, 2) enter to edit, 3) minus to exclude, and 4) backspace to delete.}
    \label{fig:patchview_hover}
\end{figure}

\subsubsection{Creating and Configuring View (I1)}
To make sense of this proliferation of world elements, \persona{} decides to use the view module to organize them. In \sys{}, a \emph{view} is a single GD\&M visual-semantic space that organizes world elements in relation to a specific set of user-defined concepts. \persona{} can create a new view by first clicking the View button in the bottom left corner and then clicking the + button. The user can set the concept associated with each magnet in the view either by dropping existing notes into placeholder magnets (\added{i.e., using elements as concepts}, Figure~\ref{fig:patchview_concept_add}a) or clicking the ``Type in a new magnet’’ button that shows up when the user hovers their mouse close to the placeholder or existing concepts (Figure~\ref{fig:patchview_concept_add}b). Once \persona{} configures the view with a set of concepts, she adds relevant elements \added{as dust particles in}\deleted{to} the view by dragging and dropping elements from the list module to the target view. As \persona{} adds an element to the view, \sys{} calculates its position within the view visualization space. \persona{} can also add multiple elements by first checking multiple of those in the list module. \added{Note that \persona{} can add elements of different types in a single view, if they are relevant (e.g., putting a good character and a good faction under ``good''-``evil'' view).} For concepts and elements in the view, \persona{} can read their full descriptions by hovering the cursor over them (Figure~\ref{fig:patchview_hover}).
\added{When \persona{} selects added elements from the list module, to let her know where they are in the view, the tool highlights them on the visualization.}

\subsubsection{Correcting View Visualization (I3)}
For some elements added to the view, \persona{} does not agree with how \sys{} positioned them. If \persona{} thinks the description of a particular concept is not detailed enough for the tool to grasp, she can modify it via the list module. Alternatively, she can edit the concept's definition text directly within the view module by hitting enter while hovering the cursor over the concept's magnet (similar to Figure~\ref{fig:patchview_hover}, but with concepts). As \persona{} updates the concept, \sys{} tries to reposition elements in relation to the concept. For elements still misplaced from \persona{}’s perspective, \persona{} can manually adjust their positions by dragging them in the view (Figure~\ref{fig:teaser}c1). When positioning future elements, \sys{} leverages user-adjusted elements as examples to better follow the user's perspective.

\subsubsection{Sensemaking Multiple Views (I1)}

\begin{figure}
    \includegraphics[width=0.35\textwidth]{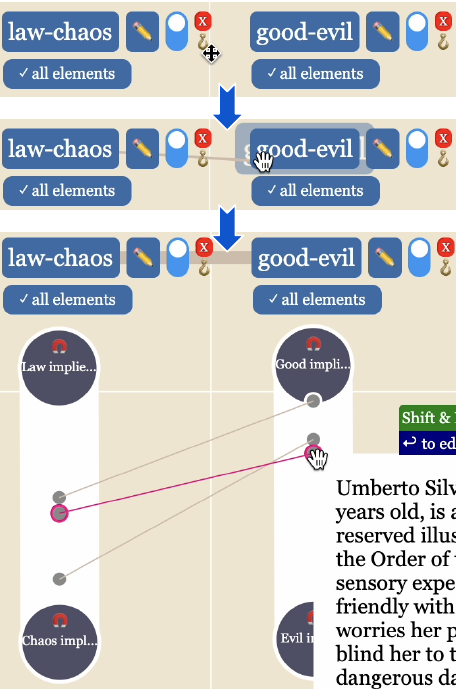}
    \caption{By dragging and dropping the anchor, the user can connect multiple views to have a better understanding of how elements distribute along those views. The same element is connected by the thin line, and for the highlighted element, the connecting line is also highlighted.}
    \Description{The figure shows the interaction of connecting multiple views together. The user drags an anchor widget of one view to another view's name. With this interaction, two views are connected, with the same element in both views connected by a thin line. As the user hovers over their mouse over one element, the same element in another view and the connecting line are highlighted in red.}
    \label{fig:patchview_anchoring}
\end{figure}

\begin{figure}
    \includegraphics[width=0.35\textwidth]{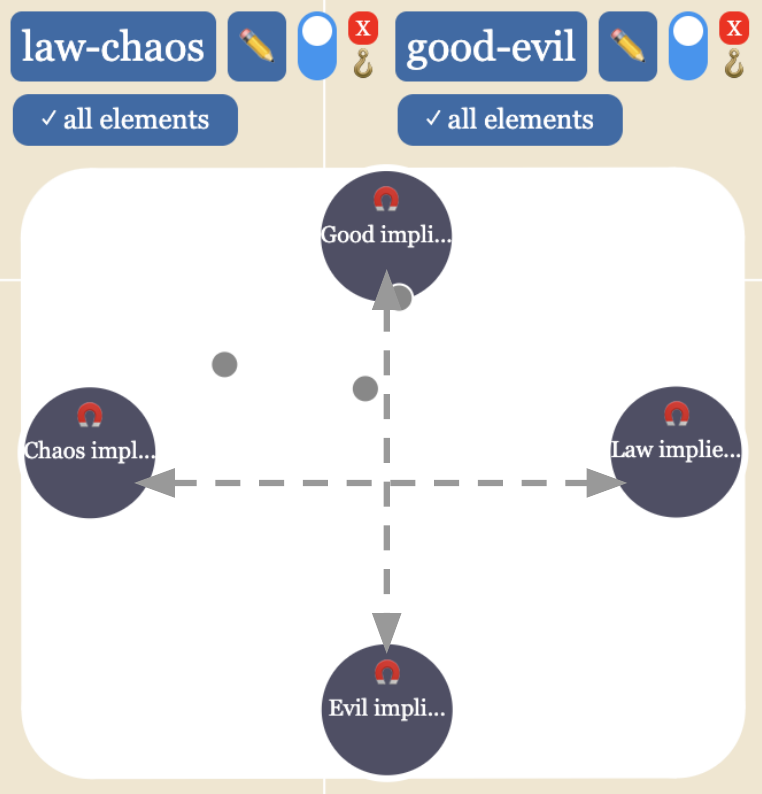}
    \caption{When two separate views are each defined by exactly two concepts, the user can cross these views into a 2D plane visualization. Analogically, this would correspond to putting dust particle elements under the simultaneous influence of two uniform magnet fields of concepts.}
    \Description{Two views that have two concepts respectively are crossed with each other, forming a visualization with two conceptual axes.}
    \label{fig:patchview_2d_plane}
\end{figure}

\begin{figure}
    \includegraphics[width=0.478\textwidth]{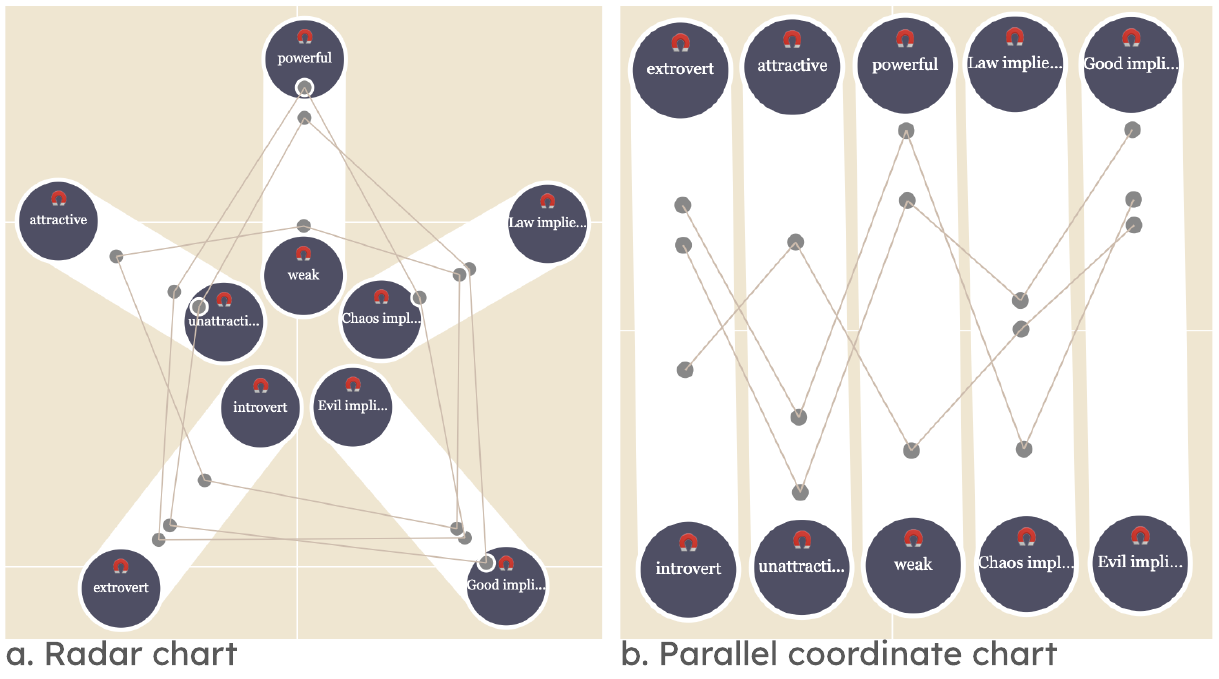}
    \caption{Multiple views can be flexibly organized to form a) a radar chart or b) a parallel coordinate chart.}
    \Description{The first subfigure shows five views with two concepts connected to each other and arranged to form a radar chart. The second subfigure shows five views with two concepts connected to each other and arranged to form a parallel coordinate chart.}
    \label{fig:patchview_radar_parallel}
\end{figure}

\persona{} continues organizing world elements by creating multiple views. \persona{} organizes these views by dragging view names and concepts. At a certain point, \persona{} realizes that it is difficult to understand how characters are distributed along the conceptual dimensions of two views, good-evil and lawful-chaotic alignments. To have a better understanding, \persona{} anchors them together and \sys{} connects the same elements in both views with a thin line (Figure~\ref{fig:patchview_anchoring}). \added{Note that only elements that exist in both views get connected.} As \persona{} hovers her cursor over an element in one of the views, the identical element in another view and the thin connecting line between these elements are highlighted (Figure~\ref{fig:patchview_anchoring}). After connecting these views, as each view is defined by only two concepts, \persona{} thinks that it would be easiest to make sense of these elements via a 2-dimensional visualization with two axes. For that, \persona{} can cross two views, and \sys{} renders the view in 2D plane visualization instead of connecting elements with lines (Figure~\ref{fig:patchview_2d_plane}). \added{Note that \sys{} only visualizes elements that exist in both crossed views.} As \persona{} adds more views, she continues to experiment with other visual arrangements, such as radar charts and parallel coordinate charts~\cite{munzner2014visualization, chung2021beyond} (Figure~\ref{fig:patchview_radar_parallel}). 

\subsubsection{Steering Generation in the View (I2)}
As \persona{} organizes world elements in the view, she finds herself wanting to add more characters to populate empty spaces within view visualizations. To steer the generation with this nuanced intention, \persona{} leverages a generative steering function on each view. \persona{} first clicks on the ``with Steering'' toggle switch at the top right to enter the generation mode. Then, \persona{} places the generation control right on the view space itself, indicating that \persona{} wants the newly generated element to be placed near the specified position (Figure~\ref{fig:teaser}b). To steer generation along multiple aspects, \persona{} can also place multiple generation controls on multiple views. After placing controls, \persona{} clicks on one of the type buttons or prompts the LLM \added{(thin and thick dashed lines in Figure~\ref{fig:patchview_gen-path}, respectively)} to generate an element with the steering constraints applied. 
After generating the element, \sys{} calculate its position in the view to place it on the view.
Similar to how \sys{} visualizes elements in the view, \sys{} leverages elements added and edited by \persona{} to adjust its steering behavior to the user's perception of concepts. 

\subsubsection{Correcting Generation (I3)}
Sometimes, generated items do not perfectly align with \persona{}’s specifications. To iterate on those, she can directly modify the text of the element or ask \sys{} to rewrite it by dragging the element while holding the shift key (Figure~\ref{fig:teaser}c2). If \persona{} disagrees with how \sys{} places a generated element, she can reposition it by dragging (Figure~\ref{fig:teaser}c1). For elements that \persona{} does not want to keep, \persona{} can either remove them from the view by hitting the minus key while hovering the cursor over the element or delete them from the view and the list by hitting the backspace key while hovering the cursor. \persona{} continues generating, editing, and organizing elements until she is done.

\subsection{Technical Details and Implementation}

We built \sys{} as a web application with a React-based frontend and a Node-based backend server. We provide technical details on 1) mapping between the position of the element and the its relevance to the concepts and 2) LLM prompting.

\subsubsection{Mapping Between Position and Weight}

To enable visualization and visual steering on the view, \sys{} needs to map the visualized position of an element to its relevance to considered concepts and vice versa. 
Here, we quantified the relevance of an element to the concept as \emph{weight} values between 0 and 1.
We first forced the placement of concepts to be convex, as the non-convex arrangement of concepts can bring in more complexities with those mappings. With the convex arrangement of concepts, we can compute the element position easily by weight-summing the concept positions with weights on those concepts. However, deciding weights from the element position is not trivial if there are more than three concepts, as a single position does not fall into one weight combination. That is, with more than three concepts, there can be more than three weights that need to be decided, but there would be only three equations with a 2D arrangement of concepts: 
\begin{equation}
    \begin{aligned}
        \Sigma_{i=1}^n w_i x_i = x_e \\ 
        \Sigma_{i=1}^n w_i y_i = y_e \\
        \Sigma_{i=1}^n w_i = 1
    \end{aligned}
    \label{eq:weight}
\end{equation}

$x_i$ and $y_i$ stand for the position of each concept, while $x_e$ and $y_e$ indicate the position of the element. $w_i$ stands for the weight that needs to be inferred and $n$ is the number of concepts.

Due to the above reason, with more than three concepts, we used the following heuristic to compute one weight combination: With all combinations of three concepts from all concepts, we first calculated weights for each combination. Then, we filtered out combinations with negative weights. After that, we calculated a weight for each concept by summing all weights from all the left combinations. Then, we finalized the weights by dividing each weight by the sum of all weights for all concepts. 
Note that with this approach, steering element generation with more than three concepts does have a limitation as not all possible weights are expressible with one geometric positioning of concepts. When two axes are crossed to form a 2D plane visualization (as in Figure~\ref{fig:patchview_2d_plane}), for each view, we first calculated the crossing point of the following two lines: 1) the line that passes through two concepts of the view and 2) the line that passes through the position of the element and is parallel to the line of two concepts from the other view. Then, as this calculated point is on the line that passes through two concepts of the view, we can calculate the weight from the equation~\ref{eq:weight}.

\subsubsection{LLM Prompting}
\begin{figure*}
    \includegraphics[width=\textwidth]{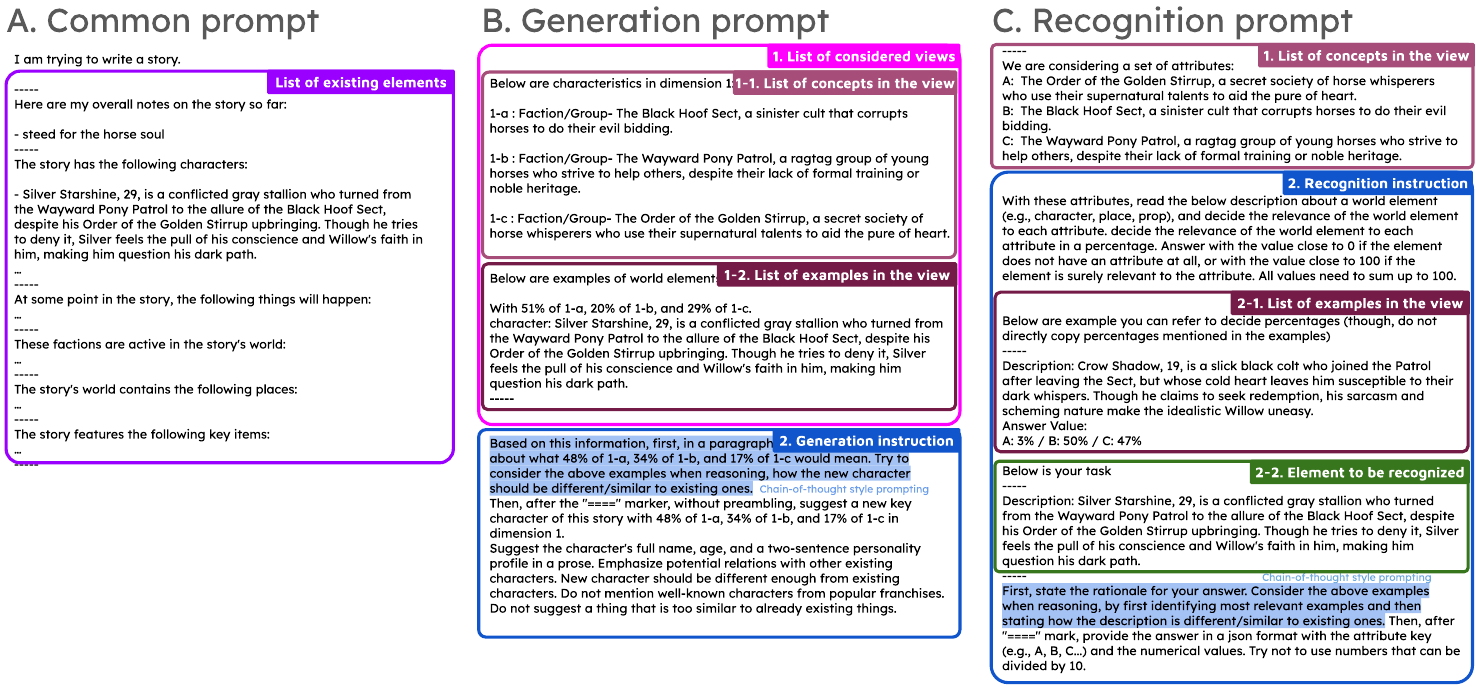}
    \caption{Prompts used for \sys{}.}
    \Description{Prompts used for large language models are given. The first prompt is a common prompt, which starts with "I am trying to write a story." It continues with a block saying "list of existing elements," which is saying "Here are my overall notes on the story so far: -steed for the horse soul. / The story has the following characters: - Silver Starshine, 29, is a conflicted gray stallion who turned from the Wayward Pony Patrol to the allure of the Black Hoof Sect, despite his Order of the Golden Stirrup upbringing. Though he tries to deny it, Silver feels the pull of his conscience and Willow's faith in him, making him question his dark path. / At some point in the story, the following things will happen: ... / These factions are active in the story's world: ... / The story's world contains the following places: ... / The story features the following key items: ...". The second prompt is a generation prompt, with the first block "list of considered views" which has two sub-blocks. The first sub-block is "list of concepts in the view," saying "Below are characteristics in dimension 1: 1-a: Faction/Group- The Black Hoof Sect, a sinister cult that corrupts horses to do their evil bidding. 1-b: Faction/Group- The Wayward Pony Patrol, a ragtag group of young horses who strive to help others, despite their lack of formal training or noble heritage. 1-c: Faction/Group- The Order of the Golden Stirrup, a secret society of horse whisperers who use their supernatural talents to aid the pure of heart." The second sub-block is "list of examples in the view" saying "Below are examples of world elements with varying values in dimension 1. With 51\% of 1-a, 20\% of 1-b, and 29\% of 1-c. Character: Silver Starshine, 29, is a conflicted gray stallion who turned from the Wayward Pony Patrol to the allure of the Black Hoof Sect, despite his Order of the Golden Stirrup upbringing. Though he tries to deny it, Silver feels the pull of his conscience and Willow's faith in him, making him question his dark path." The generation instruction block follows, which is saying "Based on this information, first, in a paragraph, in dimension 1, reason about what 48\% of 1-a, 34\% of 1-b, and 17\% of 1-c would mean. Try to consider the above examples when reasoning, how the new character should be different/similar to existing ones. Then, after the "====" marker, without preambling, suggest a new key character of this story with 48\% of 1-a, 34\% of 1-b, and 17\% of 1-c in dimension 1. Suggest the character's full name, age, and a two-sentence personality profile in a prose. Emphasize potential relations with other existing characters. New character should be different enough from existing characters. Do not mention well-known characters from popular franchises. Do not suggest a thing that is too similar to already existing things." The initial part of generation instruction is annotated as chain-of-thought style prompting. The third prompt is recognition prompt, with the first block of "list of concepts in the view." The block is saying "We are considering a set of attributes: A:  The Order of the Golden Stirrup, a secret society of horse whisperers who use their supernatural talents to aid the pure of heart. B:  The Black Hoof Sect, a sinister cult that corrupts horses to do their evil bidding. C:  The Wayward Pony Patrol, a ragtag group of young horses who strive to help others, despite their lack of formal training or noble heritage." The second block follows, which is entitled recognition instruction. It starts with the following: "With these attributes, read the below description about a world element (e.g., character, place, prop), and decide the relevance of the world element to each attribute. decide the relevance of the world element to each attribute in a percentage. Answer with the value close to 0 if the element does not have an attribute at all, or with the value close to 100 if the element is surely relevant to the attribute. All values need to sum up to 100." Then, list of examples in the view follows as a sub-block: "Below are example you can refer to decide percentages (though, do not directly copy percentages mentioned in the examples) / Description: Crow Shadow, 19, is a slick black colt who joined the Patrol after leaving the Sect, but whose cold heart leaves him susceptible to their dark whispers. Though he claims to seek redemption, his sarcasm and scheming nature make the idealistic Willow uneasy. / Answer Value: A: 3\% / B: 50\% / C: 47\%". "Element to be recognized" follows as another sub-block: "Below is your task / Description: Silver Starshine, 29, is a conflicted gray stallion who turned from the Wayward Pony Patrol to the allure of the Black Hoof Sect, despite his Order of the Golden Stirrup upbringing. Though he tries to deny it, Silver feels the pull of his conscience and Willow's faith in him, making him question his dark path." The recognition instruction ends with "First, state the rationale for your answer. Consider the above examples when reasoning, by first identifying most relevant examples and then stating how the description is different/similar to existing ones. Then, after "====" mark, provide the answer in a json format with the attribute key (e.g., A, B, C...) and the numerical values. Try not to use numbers that can be divided by 10." The initial part of this prompt is annotated as "chain-of-thought style prompting."}
    \label{fig:patchview_prompt}
\end{figure*}

To generate elements with steering inputs and recognize the relevance of elements to concepts, we prompted \texttt{claude-2.0} and \texttt{claude-instant-1.2} from Anthropic~\cite{claude2}, respectively.
We chose these models because they have shown better performance in creative writing contexts than leading alternatives~\cite{chakrabarty2024art}.

Prompts for both generation and recognition began by introducing a set of existing world elements for context, as in Figure~\ref{fig:patchview_prompt}a. By default, all existing world elements were supplied as part of this context; the user could also select a subset of existing elements to ensure that only those elements would be provided as context.

When generating new world elements without visual steering input, introductory context was followed directly by an instruction describing what kind of element to generate.
When generating with visual steering input, we first appended concepts of all views \added{(Figure~\ref{fig:patchview_prompt}b-1-1)} and examples of how existing elements have relevance to those concepts \added{(Figure~\ref{fig:patchview_prompt}b-1-2)}. 
These examples came from elements that the user has already placed in the view, including those repositioned by the user. 
Note that in the prompt, all views and concepts are phrased as ``dimensions’’ and ``characteristics’’, respectively.
A chain-of-thought~\cite{wei2022chain} style generation instruction prompt followed after \added{(Figure~\ref{fig:patchview_prompt}b-2)}, which asked the LLM to first reason about how the element description should be written considering steering inputs and then to write the element description.

Recognition of concept relevance values takes place on a per-view basis, so a prompt for the recognition task included concept descriptions for only a particular considered view, as in Figure~\ref{fig:patchview_prompt}c. In the recognition prompts, introductory context \added{(Figure~\ref{fig:patchview_prompt}a)} and concept descriptions \added{(Figure~\ref{fig:patchview_prompt}c-1)} were followed by instructions about how to interpret numbers \added{(Figure~\ref{fig:patchview_prompt}c-2)}, and then by examples of the correct performance of a recognition task for this set of concepts \added{(Figure~\ref{fig:patchview_prompt}c-2-1)}.
Because these examples were taken from past placements of elements into this specific view, information about how the user repositioned elements in this view were taken into account at this step.
Finally, the world element to be analyzed was attached \added{(Figure~\ref{fig:patchview_prompt}c-2-2)}, with the chain-of-thought~\cite{wei2022chain} instruction that the LLM should provide reasoning before the result. The LLM was asked to provide recognition results in a JSON format with concept identifiers as keys and concept relevance weights as values.
\section{User Study}

We conducted a user study on \sys{} to learn if it supports sensemaking and steering of world element generation under the user’s unique story world context. Specifically, we tried to answer the following research questions to determine if \sys{} effectively supports the interactions described in Section~\ref{sec:interaction_alignment}.

\begin{itemize}
    \item RQ1: Does \sys{} help the user with sensemaking world elements? (I1)
    \item RQ2: Does \sys{} help users express nuanced intentions with visual steering? (I2)
    \item RQ3: Does \sys{} help users correct AI results and behaviors? (I3)
\end{itemize}

Additionally, we aimed to discover how \sys{} might be used in the worldbuilding process. 
\begin{itemize}
    \item RQ4: How do users leverage features of \sys{} for worldbuilding?
\end{itemize}

To answer these questions, we conducted a study that mixes a within-subject comparative task and an observational task, along with both quantitative and qualitative analyses of collected data. 

\subsection{Participants}

\begin{table}[]
\caption{Participant backgrounds. AI Exp* stands for experience with generative AI technologies (e.g., LLM, text-to-image models), the former denoting any experience and the latter indicating the use in their writing practice.}
\begin{tabular}{lllll}
\hline
   & Expertise & Year & Domain                    & AI Exp* \\ \hline
P1 & Hobbyist  & 10   & novel                     & Y/Y    \\
P2 & Hobbyist  & 19   & novel, TRPG               & N/N    \\
P3 & Hobbyist  & 8    & novel                     & Y/N    \\
P4 & Hobbyist  & 6    & novel                     & Y/Y    \\
P5 & Hobbyist  & 7    & novel                     & N/N    \\
P6 & Hobbyist  & 5    & novel                     & Y/Y    \\
P7 & Expert    & 8    & screenwriting, game, TRPG & Y/Y    \\
P8 & Hobbyist  & 25   & novel, fan fiction        & N/N    \\
P9 & Hobbyist  & 5    & novel                     & Y/Y   \\ \hline
\end{tabular}
\Description{This table presents the backgrounds of nine participants. The first row contains participant P1, who is a hobbyist with 10 years of experience in writing novels and has experience with generative AI technologies, including using them in their writing practice. The second row describes participant P2, a hobbyist with 19 years of experience in writing novels and tabletop role-playing games (TRPG), but has no experience with generative AI technologies. Participant P3, in the third row, is a hobbyist with 8 years of experience in writing novels and has experience with generative AI technologies, but does not use them in their writing practice. The fourth row introduces participant P4, a hobbyist with 6 years of experience in writing novels and has experience with generative AI technologies, including using them in their writing practice. Participant P5, in the fifth row, is a hobbyist with 7 years of experience in writing novels and has no experience with generative AI technologies. The sixth row presents participant P6, a hobbyist with 5 years of experience in writing novels and has experience with generative AI technologies, including using them in their writing practice. Participant P7, in the seventh row, is an expert with 8 years of experience in screenwriting, game writing, and tabletop role-playing games (TRPG), and has experience with generative AI technologies, including using them in their writing practice. The eighth row describes participant P8, a hobbyist with 25 years of experience in writing novels and fan fiction, but has no experience with generative AI technologies. Finally, the ninth row introduces participant P9, a hobbyist with 5 years of experience in writing novels and has experience with generative AI technologies, including using them in their writing practice.}
\label{tab:participants}
\end{table}

We recruited nine participants (four women, three men, one nonbinary, and one who did not disclose gender, ages 24-51, M=33.4, SD=8.7) through Upwork\footnote{https://www.upwork.com/}, a gigwork platform. We focused on recruiting hobbyists with extensive years of experience (at least five) or professionals who make a living out of story writing and worldbuilding. Participants were proficient in English. Six participants had experience using AI for story writing, and among them, five actively used AI for their practice. We detail participants in Table~\ref{tab:participants}.

\subsection{Procedure}
The study was conducted remotely via Google Meet\footnote{https://meet.google.com/}. After welcoming the participants, we asked if they were okay with recording the session. Then, we asked participants to watch two instruction videos, each on 1) the overview of \sys{} and ways to generate or create notes on the list module and 2) reading view visualizations. After each video, participants were given an opportunity to experiment with the functions that had just been introduced.

After two instruction videos, we asked participants to conduct the first task, answering sensemaking questions (RQ1). Specifically, we provided two types of questions: 1) \emph{landscape questions}, characterizing the distribution of world elements in relation to specific concepts (e.g., To which faction most characters are associated with?), and 2) \emph{comparison questions}, comparing different characters according to their relevance to concepts (e.g., Which character is most associated with faction A?). \added{These were multiple choice questions with one correct option.} We measured whether the participants were correct and the time taken to answer. We expected that if the visualization could help users with sensemaking, they would answer more accurately in less time.

Participants conducted the task in a within-subject fashion, in two conditions: only with the list interface of Figure~\ref{fig:patchview_interface}b (baseline) and together with the view visualization (treatment). We prepared two collections of elements, both focusing on character descriptions. One collection considered three different factions to characterize elements. Another considered two axes of good-evil and law-chaos as concepts, which are often used as character alignment structures for role-playing games such as \emph{Dungeons \& Dragons}~\cite{livingstone1986dicing}. We populated each collection with 10 characters generated with \sys{}. To visualize characters, we leveraged \sys{}'s recognition results. The authors crafted questions after carefully reading through all generated elements (Appendix~\ref{sec:appendix_sensemaking}). For each collection, we asked participants to answer both types of questions, with one collection given the baseline condition and the other with the treatment condition. We randomized the order of conditions to minimize ordering effects.  
For each question, we asked participants first to open the link to the question. After they understood the question, we asked them to open the link to the tool and answer the question with the story world provided in the tool. We timed the time taken to answer questions. 

After the first task, participants watched three more videos on \sys{}’s functions: 1) creating and configuring views, 2) generating and editing world elements with views, and 3) rewriting elements and connecting multiple views. As before, participants were allowed to experiment with the just-introduced functions after finishing each video. 

Once all functions were introduced, we asked participants to perform a second task: building a story world with \sys{} while thinking aloud. We asked them to create at least one view and put five elements in the view. Moreover, we asked participants to place elements in the view where they think should be when finishing the task. 
Through this task, we wanted to understand if participants could use \sys{} with concepts of their interest, visualizing elements (RQ1), steering generation with their nuanced intentions (RQ2), and correcting AI results and behaviors during the usage (RQ3). Moreover, we wanted to learn how \sys{} supports the worldbuilding process (RQ4).

To understand participant behavior during this task, we collected logs of \sys{} usage, including concepts that participants considered, steering inputs they made, outputs they received from \sys{}, and corrections that they made to outputs. We also collected screen and think-aloud recordings. Participants could spend at most 40 minutes on this task. After the task, we asked participants to complete a survey and an interview. The survey asked about the helpfulness of each \sys{} feature and included Creativity Support Index~\cite{cherry2014quantifying} questions on enjoyment, exploration, expressiveness, immersion, and the results of tool usage being worth the effort. Note that we did not use Creativity Support Index questions to compare the tool to others, only to gather participants' overall impressions of the tool. The interview aimed to elicit detailed perceptions about functions of \sys{} and how \sys{} could be used in participants' actual practices. The whole study took at most 120 minutes. Each participant received \$60 for participation.

\subsection{Results}

\begin{figure*}[t]
    \includegraphics[width=\textwidth]{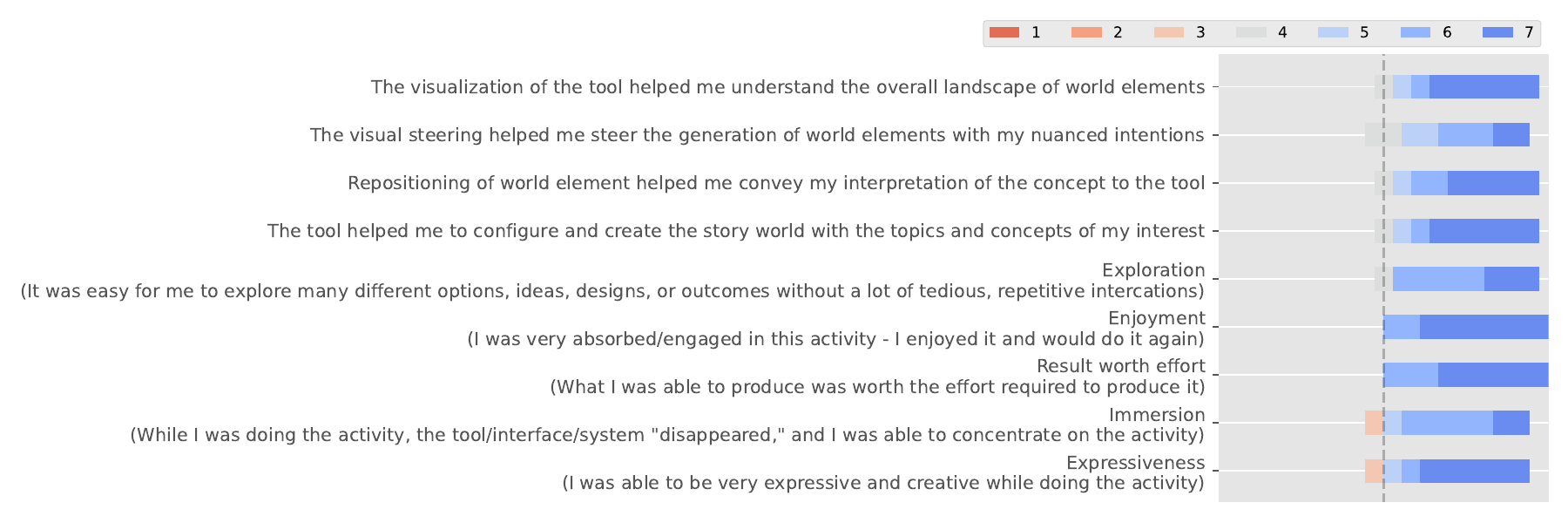}
    \caption{Survey results.}
    \Description{Survey responses about the helpfulness of a visualization tool on a 7-point Likert scale from 1 (Strongly Disagree) to 7 (Strongly Agree). Statements rated include: "The visualization of the tool helped me understand the overall landscape of world elements", "The visual steering helped me steer the generation of world elements with my nuanced intentions", "Repositioning of world element helped me convey my interpretation of the concept to the tool", "The tool helped me to configure and create the story world with the topics and concepts of my interest", "Exploration - It was easy for me to explore many different options, ideas, designs, or outcomes without a lot of tedious, repetitive interactions", "Enjoyment - I was very absorbed/engaged in this activity - I enjoyed it and would do it again", "Result worth effort - What I was able to produce was worth the effort required to produce it", "Immersion - While I was doing the activity, the tool/interface/system 'disappeared,' and I was able to concentrate on the activity", and "Expressiveness - I was able to be very expressive and creative while doing the activity". Responses skew positive, with most statements receiving an average rating between 5-7 on the scale. Only for Immersion and Expressiveness, there was one respondent who answered negatively, for each respectively.}
    \label{fig:patchview_survey}
\end{figure*}

\begin{figure}
    \includegraphics[width=0.478\textwidth]{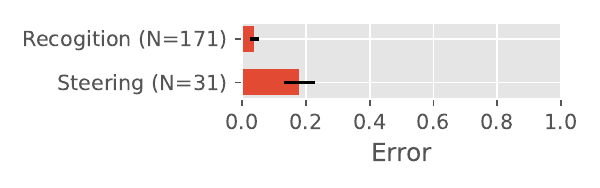}
    \caption{Errors in 1) recognizing the concept weights for elements placed in the visualization and 2) steering the generation of elements according to concept weights specified by the user, measured on a 0-to-1 scale. The error bars in this paper indicate the 95\% confidence intervals.}
    \Description{The figure presents a bar chart showing the errors for two tasks, Recognition and Steering, measured on a scale from 0 to 1. The Recognition task, with a sample size of 171, has a very small error close to 0. The Steering task, with a sample size of 31, shows an error of 0.18, with an error bar of width around 0.08, which indicates the confidence interval.}
    \label{fig:patchview_error}
\end{figure}

\begin{figure}
    \includegraphics[width=0.478\textwidth]{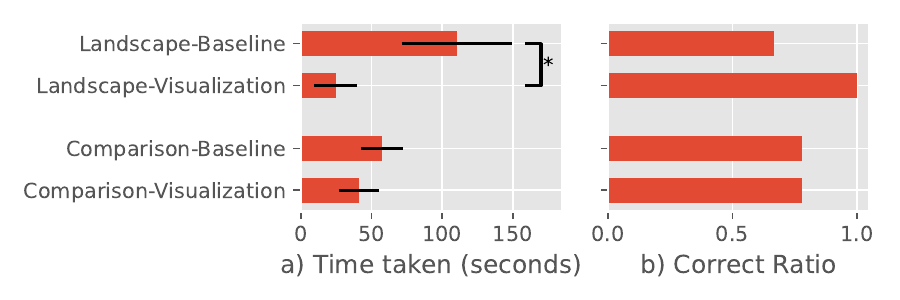}
    \caption{Sensemaking question results. a) Time taken for answering questions and b) correct ratio. * indicates $p<0.001$.}
    \Description{Bar graph showing sensemaking question results from Baseline and Visualization conditions. Specifically, the results of landscape questions are compared separately from those of comparison questions. Graph (a) depicts the average time taken in seconds to answer questions for each condition. For landscape questions, the Baseline condition took around 110 seconds, which is significantly longer than the Visualization condition, which took around 25 seconds. The significance is noted with asterisks. For comparison questions, the Baseline condition took slightly longer than the Visualization condition, but the difference is not significant. Graph (b) shows the ratio of correct responses for each condition. For landscape questions, the baseline condition shows 0.67 of correctness, while the visualization condition shows 1.0 of correctness. For the comparison question, both baseline and visualization conditions have the same correct ratio of around 0.78.}
    \label{fig:patchview_sensemaking}
\end{figure}

We analyzed survey responses (Figure~\ref{fig:patchview_survey}), recognition and steering errors from log data (Figure~\ref{fig:patchview_error} and \ref{fig:patchview_example_incontext}), answer time and correctness of the sensemaking questions (Figure~\ref{fig:patchview_sensemaking}), video recordings, and interview data. 
We measured recognition errors by the difference between \added{\sys{}'s automatic placements of elements and the user's final placements of the same elements in views}\deleted{ where \sys{} placed elements and where the user placed them in views}. 
\added{This error will be zero if the user does not reposition the placement, and one if the user repositions an extreme value to another extreme one (e.g., fully good to fully evil).}
We calculated steering errors by the difference between where the user placed steering inputs and the user's \added{final} placement of world elements generated \added{with}\deleted{in response to} these inputs. 
\added{This error will be zero if the content of the generated element perfectly aligns with the user input, and one when the AI generates a totally misaligned element with the user input (e.g., a fully evil character generated with fully good input).}
This approach measures errors from the natural usage of the tool. 
However, \added{note that this approach also has a limitation, as correcting the error does have the cost of moving the element. Moreover,}\deleted{it has a limitation in that} it cannot consider errors of elements deleted during usage, as it requires the final placement of the element in the view by the user.
Note that participants did not delete a high number of elements---participants deleted 10 elements out of 181 for recognition and two elements out of 33 for steering. 
Moreover, we could not collect errors for rewriting interactions due to a technical issue. We analyzed video recordings and interview data by iterative coding with inductive analysis. 

\subsubsection{RQ1: Visualization helped users with sensemaking world elements.} 

The participants \added{seemed to} largely agree with how \sys{} placed world elements in the view. Figure~\ref{fig:patchview_error} shows that the mean recognition error was measured to be 0.04 on a 0-to-1 scale for the user's arbitrary concepts. 
This result resonates with participants' interview responses ($N=6$). For instance, P9 mentioned that \sys{} accurately recognized concept relevance even in challenging cases: \inquote{It actually grasped my intention even though I gave two words, basically.}

With largely accurate automatic visualization, in the first survey question (Figure~\ref{fig:patchview_survey}), participants responded that \sys{} helped them understand the landscape of elements in the story world. The helpfulness of visualization also manifests in the sensemaking question results (Figure~\ref{fig:patchview_sensemaking}), specifically for landscape questions. When answering landscape questions, participants were significantly faster with visualization than without (Mann-Whitney $U=79$, $n_1=n_2=9$, $p<0.001$\footnote{We used non-parametric test due to small sample size and non-equal variances.}) and more correctly answered questions. However, for comparison questions, there was no significant difference in time taken to answer questions between conditions (Mann-Whitney $U=60$, $n_1=n_2=9$, $p>0.05$). Moreover, participants were similarly accurate in answering comparison questions. 

Interview results resonated with these findings: participants mentioned that they could easily understand the landscape of world elements with the help of visualization ($N=9$), allowing them to track generated elements while keeping the world under the rule and the structure. P1 mentioned: \inquote{The different views and stuff, actually seeing that on there and keeping track of it, I think, would be helpful. ... Because I end up building up too many and then I forget what the differences in each one's personality are sometimes.} 
P9 also appreciated the customizability of the visualization. 

\sys{}'s visualization also influenced how participants thought about each concept. That is, when participants do not agree with \sys{}'s placement of elements, some participants reflected on their own perception of concepts ($N=5$), often concretizing how they think about concepts. For example, P4 mentioned: \inquote{When I put this `strength', I was, kind of, just thinking about overall strength. And then, it interpreted it as physical strength. So I was like, `this is now all across these (elements), and now that really gives a more nuanced idea of… power'} (Figure~\ref{fig:patchview_study_example}). However, there were also cases where the participant had a strong idea of how they think about concepts, and for those cases, they realized that they would need to sharpen their verbiage about the concept ($N=3$). Some mentioned that future versions of the tool can explicitly support it. For example, P3 mentioned: \inquote{I think if there's a pop-up that says, `Can you give more information' or something like that, I think that would help me to force some clarity before it visualizes.}

\subsubsection{RQ2: Visual steering of \sys{} allowed users to steer the generation with nuanced intentions.}

The results indicate that the steering function was fairly accurate when used for arbitrary concepts of the participants' interests. The mean steering error was measured to be 0.18 on a 0-to-1 scale (Figure~\ref{fig:patchview_error}). As the ordinal scale of five on a bi-directional dimension is often considered to be easily discernible by people~\cite{socher-etal-2013-recursive}, if we assume uniform intervals between levels (which is often used in ML~\cite{socher-etal-2013-recursive, zhang2017predicting}), the error of 0.18 would be smaller than a single gap in a five-level ordinal scale (0.25). Hence, we conclude that \sys{} allows users to steer the generation accurately in a granularity finer than easily discernible five-level scale on dimensions with two concepts. While this standard would need to be different for cases when a view has more than two concepts, in our study, only four steered generation results considered more than two concepts. 
As in the second question in Figure~\ref{fig:patchview_survey}, participants also perceived that the tool helped them steer the world element generation with their nuanced intentions. 

Participants mentioned that visual steering for element generation and rewriting was intuitive ($N=7$). For example, for visual rewriting interaction, P6 mentioned: \inquote{All I had to do is to move where I wanted the story element to be reconnected to and that's like a no-brainer that just takes a couple of mouse clicks and you're good to go.} Participants also noted that visual steering helped them express nuanced intentions, even allowing them to realize the semantic space that they could not think about ($N=6$). For example, P2 mentioned that they could use visual steering to create a set of characters that would make more conflicts than randomly generating them. On the other hand, participants thought that natural language prompts often require more cognitive effort as they need to bring up specific instructions ($N=2$). However, participants thought that natural language prompts are beneficial as the user can be more specific in the instruction ($N=4$). With different strengths, some participants ($N=2$) thought that visual steering and natural language prompting complement each other, as P7 mentioned, by \inquote{choosing the point via steering and then giving it a little bit of (natural language) input.} For example, one limitation of the current visual rewriting interaction is that it often changes aspects the user likes. Adding a natural language prompt, such as specifying which aspects not to change, could have alleviated the issue.

\begin{figure}
    \includegraphics[width=0.478\textwidth]{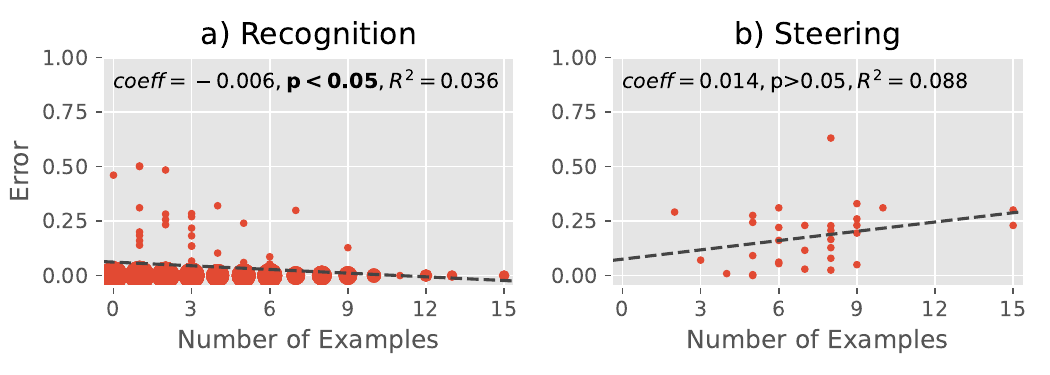}
    \caption{a) Recognition and b) steering errors according to the number of examples added by the user. Dot sizes indicate the number of data points with the same errors and example numbers. Each plot contains the linear regression result.}
    \Description{Scatter plots showing recognition and steering errors versus the number of user-provided examples. Plot a) displays recognition error on the y-axis from 0.0 to 1.0 against the number of examples on the x-axis ranging from 0 to 15. The scatter points, whose size indicates the number of data points at each coordinate, show recognition error generally decreasing as more examples are added. The linear regression line has a negative slope, with the equation coeff= -0.006, p<0.05, R^2=0.036. Plot b) shows steering error on the y-axis from 0.0 to 1.0 versus the number of examples on the x-axis from 0 to 15. The steering errors are more varied with an insignificantly fitting line of coeff=-0.014, p>0.05, R^2=0.088.}
    \label{fig:patchview_example_incontext}
\end{figure}

\subsubsection{RQ3: With more user examples, \sys{} could only improve recognition, not the steering, but to a small extent.} In the third question of Figure~\ref{fig:patchview_survey}, most participants answered that they could convey their interpretations of the concepts through repositioning elements. Similarly, during the interview, participants mentioned that it was easy to revise AI results by simply moving elements on the view or by rewriting interactions ($N=6$). For example, P7 mentioned: \inquote{I really like that you have the ability to say, kind of like, `No I'm telling you where this should go' versus `I want you to actually adjust it to fit there.'}

However, the user's correction of concepts through the addition of more examples did not turn into dramatic changes in AI behaviors. As in Figure~\ref{fig:patchview_example_incontext}a, when we conducted a linear regression on the relation between the number of examples and the recognition error, the addition of more examples significantly decreased errors ($p<0.05$), but with a small magnitude ($coeff=-0.006$) and a small ability to explain variations ($R^2=0.036$). The analysis on steering errors (Figure~\ref{fig:patchview_example_incontext}b) revealed no significance in the correlation between the number of added examples and errors ($p>0.05$). 
These resonated with the interview responses: participants felt that the study session was not long enough to sense that the tool is learning from what they are doing in the tool ($N=3$). P7 mentioned that rather than having such tool behavior changes implicit, making them more explicit to the user would be helpful: \inquote{I would have had to play with it a lot more to know if it actually was learning ... It'd be interesting if I could have a feature to refresh ... So a refresh thing would help me see what it was learning from me.}

\subsubsection{RQ4: Participants could flexibly create their own story world and suggested ways to improve the tool for more comprehensive story writing.}

\begin{table}[]
\caption{Views created by participants. $\times$ indicates that multiple views are either tied or crossed with each other. For cases with more than two concepts in view, we noted commonalities between concepts instead of directly showing the concepts themselves. El and char stand for element and character, respectively.}
\begin{tabular}{llll}
\noalign{\global\arrayrulewidth=0.4mm}
\hline
\noalign{\global\arrayrulewidth=0.15mm}
                    & View concepts                                                                                                & El \# & El Type                                                          \\ 
                    \noalign{\global\arrayrulewidth=0.4mm}
                    \hline
                    \noalign{\global\arrayrulewidth=0.15mm}
P1                  & good - evil $\times$ law - chaos                                                                                    & 9         & char                                                                  \\ \hline
P2                  & four factions                                                                                                & 9        & char                                                                  \\ \hline
P3                  & three life focuses                                                                                           & 6         & char                                                                  \\ \hline
\multirow{4}{*}{P4} & story timeline (beginning - end)                                                                             & 4         & event                                                                 \\ 
\cline{2-4} 
                    & \begin{tabular}[c]{@{}l@{}}magical aptitude (high - low) \\ $\times$ physical strength (strong - weak)\end{tabular} & 10        & char, faction                                                         \\ \cline{2-4} 
                    & good - evil                                                                                                  & 8         & char, faction                                                         \\ \hline
\multirow{2}{*}{P5} & good - evil $\times$ law - chaos                                                                                    & 3         & faction, place                                                        \\ \cline{2-4} 
                    & three locations                                                                                              & 1         & character                                                             \\ \hline
P6                  & cats - pugs $\times$ library - tombs                                                                                & 13        & \begin{tabular}[c]{@{}l@{}}char, faction, \\ prop, event\end{tabular} \\ \hline
\multirow{3}{*}{P7} & good - evil $\times$ law - chaos                                                                                    & 5         & char                                                                  \\ \cline{2-4} 
                    & story timeline (beginning - end)                                                                             & 3         & event                                                                 \\ \cline{2-4} 
                    & two factions                                                                                                 & 5         & char                                                                  \\ \hline
\multirow{3}{*}{P8} & good - evil                                                                                                  & 8         & char                                                                  \\ \cline{2-4} 
                    & law - chaos                                                                                                  & 8         & char                                                                  \\ \cline{2-4} 
                    & honest - deceitful                                                                                           & 5         & char                                                                  \\ \hline
\multirow{3}{*}{P9} & \begin{tabular}[c]{@{}l@{}}logical - emotional\\ $\times$ science-oriented - belief-oriented\end{tabular}           & 7         & char, event                                                           \\ \cline{2-4} 
                    & positive event - negative event                                                                              & 3         & event                                                                 \\ \cline{2-4} 
                    & three factions                                                                                               & 5         & char                                                                  \\ 
                    \noalign{\global\arrayrulewidth=0.4mm}
                    \hline
                    \noalign{\global\arrayrulewidth=0.15mm}
\end{tabular}
\label{tab:views_created}
\Description{Views created by participants. The table contains 9 rows, each corresponding to a participant ID (P1 to P9). For each participant, their created views are described in the "View concepts" column, with the number of elements (El #) and the element type (El Type) specified. P1's view involves the concepts of "good - evil × law - chaos", containing 9 character elements. P2's view is based on "four factions", with 9 character elements. P3's view focuses on "three life focuses", containing 6 character elements. P4's view incorporates "story timeline (beginning - end)", "magical aptitude (high - low) × physical strength (strong - weak)", and "good - evil", containing 4 events, 10 characters/factions, and 8 characters/factions, respectively. P5's view centers around "good - evil × law - chaos" and "three locations", with 3 factions/places and 1 character element, respectively. P6's view is composed of "cats - pugs × library - tombs", involving 13 characters/factions/props/events. P7's view deals with "good - evil × law - chaos", "story timeline (beginning - end)", and "two factions", using 5 characters, 3 events, and 5 characters, respectively. P8's view includes "good - evil", "law - chaos", "honest - deceitful", employing 8 characters, 8 characters, and 5 characters, respectively. Lastly, P9's view covers "logical - emotional x science-oriented - belief-oriented", "positive event - negative event" and "three factions", containing 7 characters/events, 3 events, and 5 characters.}
\end{table}

\begin{figure*}
    \includegraphics[width=\textwidth]{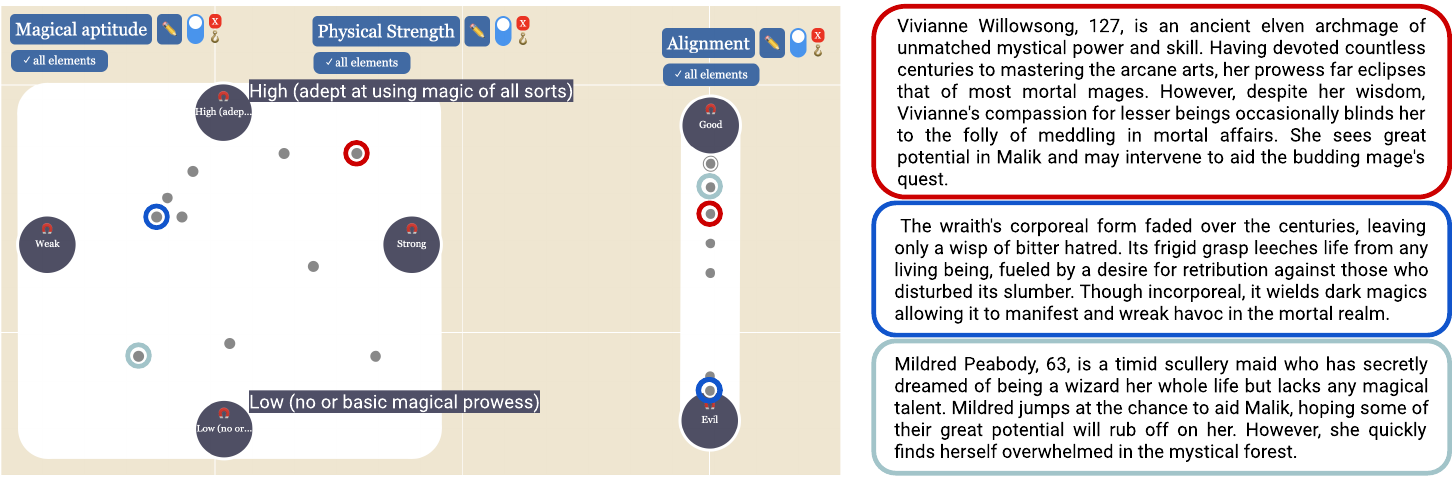}
    \caption{Views created by P4, organized by magical aptitude, physical strength (which is noted as Weak-Strong), and good-evil alignment. Views include characters and factions as elements. Example elements are presented on the right and where they are positioned in the view is marked with the circles of the same border color. }
    \Description{On the left, there are views, one with two views crossed each other. Two views have the concept of "High (adept at using magic of all sorts)"-"Low (no or basic magical prowess)" and "Strong"-"Weak." On the right of it, there is another view of "Good"-"Evil". There are three elements highlighted, with the corresponding texts on the left of the figure. The first one is highlighted in red, being positioned close to "High (adept at using magic of all sorts)", "Strong", and somewhat "Good." The element's description is "Vivianne Willowsong, 127, is an ancient elven archmage of unmatched mystical power and skill. Having devoted countless centuries to mastering the arcane arts, her prowess far eclipses that of most mortal mages. However, despite her wisdom, Vivianne's compassion for lesser beings occasionally blinds her to the folly of meddling in mortal affairs. She sees great potential in Malik and may intervene to aid the budding mage's quest." The second element is highlighted in blue, being positioned in somewhat "High (adept at using magic of all sorts)", somewhat "Weak", and very "Evil". The text description is "The wraith's corporeal form faded over the centuries, leaving only a wisp of bitter hatred. Its frigid grasp leeches life from any living being, fueled by a desire for retribution against those who disturbed its slumber. Though incorporeal, it wields dark magics allowing it to manifest and wreak havoc in the mortal realm." The third element is in light teal, being positioned in "Low (no or basic magical prowess)," "Weak," and "Good." The text is "Mildred Peabody, 63, is a timid scullery maid who has secretly dreamed of being a wizard her whole life but lacks any magical talent. Mildred jumps at the chance to aid Malik, hoping some of their great potential will rub off on her. However, she quickly finds herself overwhelmed in the mystical forest."}
    \label{fig:patchview_study_example}
\end{figure*}

\begin{figure}
    \includegraphics[width=0.478\textwidth]{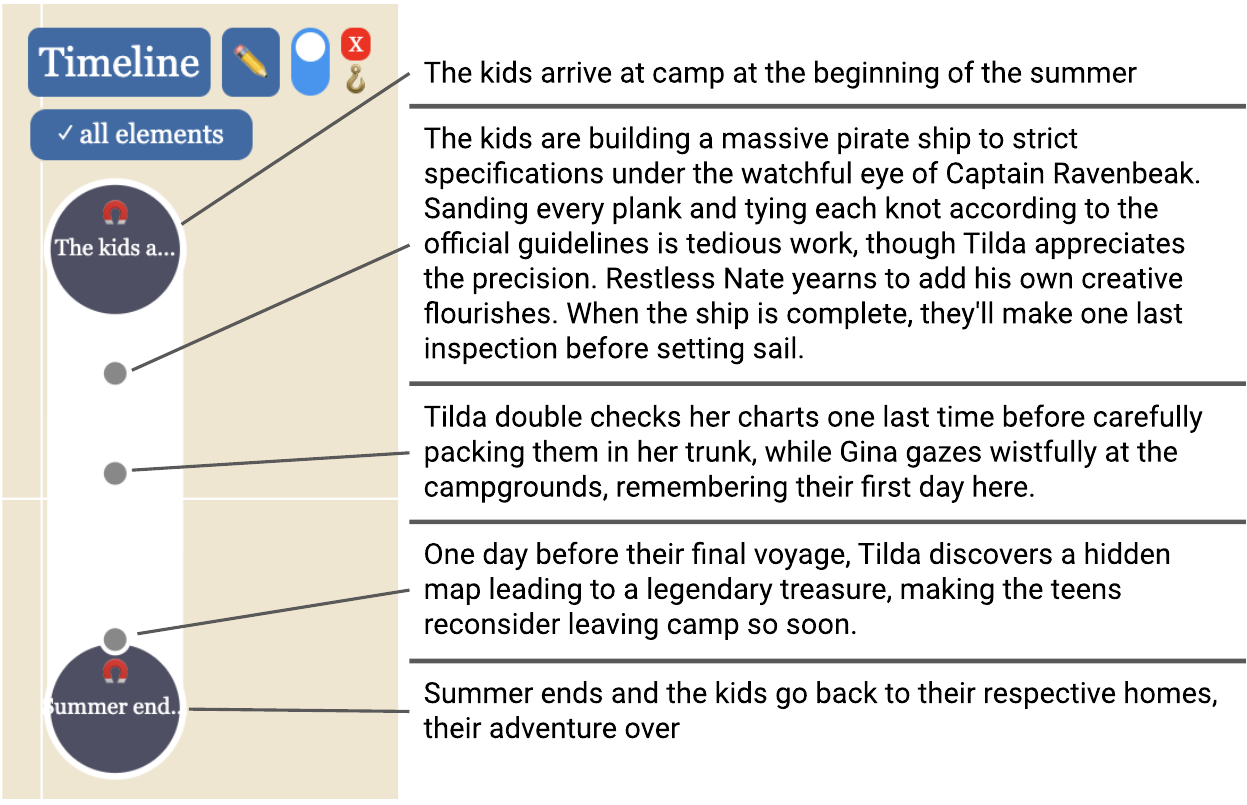}
    \caption{A timeline view created by P7.}
    \Description{There is a view named "Timeline" with two concepts as magnets. The magnet at the top says "The kids arrive at camp at the beginning of the summer," and the one at the bottom says "Summer ends and the kids go back to their respective homes, their adventure over." There are three elements in the view. From top to bottom, each is saying "The kids are building a massive pirate ship to strict specifications under the watchful eye of Captain Ravenbeak. Sanding every plank and tying each knot according to the official guidelines is tedious work, though Tilda appreciates the precision. Restless Nate yearns to add his own creative flourishes. When the ship is complete, they'll make one last inspection before setting sail," "Tilda double checks her charts one last time before carefully packing them in her trunk, while Gina gazes wistfully at the campgrounds, remembering their first day here," and "One day before their final voyage, Tilda discovers a hidden map leading to a legendary treasure, making the teens reconsider leaving camp so soon."}
    \label{fig:patchview_study_timeline}
\end{figure}

With \sys{}, participants could structure the story world according to concepts of their interest. Table~\ref{tab:views_created} shows the summary of views participants created. Many participants created views for alignments~\cite{hergenrader2018collaborative}, either good-evil ($N=5$) or law-chaos ($N=4$) dimensions. It might be because these alignments are widely used to organize characters or because we used these dimensions as examples in the tutorial. Participants also created views with custom concepts, such as factions ($N=3$), locations ($N=1$), or other concepts of the participant's interest (e.g., magical aptitude from Figure~\ref{fig:patchview_study_example}, $N=9$). One interesting view type was story timeline, where participants tried to align events between the story's beginning and end (Figure~\ref{fig:patchview_study_timeline}, $N=2$). It shows that participants would eventually want to create a coherent storyline with the world elements they created. Participants added various element types to the views, but the character was most frequently added.

As shown in the results of the fifth to ninth questions of Figure~\ref{fig:patchview_survey}, participants felt that \sys{} helped them expressively explore various ideas while enjoying and being immersed in the process, ending up with a result that was worth their efforts. Participants thought that the tool could help with ideation or filling in details for the part they are not good at ($N=8$). 
As generative features could add new things to the user's world, some participants mentioned that AI generation would be most useful for the ideation stage or the settings where the user needs to constantly come up with new elements (e.g., D\&D), rather than for the cases where the user already has a highly-structured and consistent world ($N=2$).  
P9 also mentioned that the visualization could facilitate collaboration when multiple people are working on the worldbuilding: \inquote{If you're writing, let's say with a group of people, ... it's helpful to be able to quickly see `Okay this character is in this faction`, instead of having to go through it because you might not be the one who wrote it.}

While participants appreciated \sys{}, they also made suggestions to improve the tool for worldbuilding and story writing practices. First, as story elements can have relationships with each other (e.g., relationships between characters), participants suggested features to visualize such relationships, such as arrows between elements ($N=4$). Specifically with the temporal relationships, as shown in the participants' usage patterns, participants mentioned that a view more specialized for timeline would help ($N=2$).
As some elements could change with the story's progress, P9 mentioned that it would be helpful to have the feature to ``draw out`` the trajectory of changes so that it can direct the tool to generate those changes. 
Second, as world elements could constantly change with the use of the tool, participants wanted the tool to handle the consistency between elements ($N=2$). For example, if the user edits one character's name in one note, participants wanted the tool to automatically update the character's name appearing in other notes.
Third, some participants wanted to flesh out world elements by adding relevant images with AI image generation models ($N=2$). They mentioned that additional visuals could help them not only concretize the world element but also quickly grasp it.
Lastly, some wanted a feature to import their own world to the tool if they already have the world that they are working on ($N=2$).

\section{Discussion}
We discuss \sys{}'s interaction design, utility in worldbuilding and story writing practices, technical alternatives, and limitations.

\subsection{Visually Bridging User and Generative AI}
We introduce GD\&M as a visual interaction to bridge the user and generative AIs. As mentioned in Section~\ref{sec:design}, GD\&M is most applicable when the user generates a lot of things within the conceptual dimensions of their interests. The interaction helps with the user's evaluation, specification, and alignment of AI behaviors~\cite{norman2002design, terry2023ai}. We believe this interaction can be adopted to other use cases. In the text domain, for instance, GD\&M could be used for organizing and steering idea generation~\cite{siangliulue2016ideahound} with LLMs. Beyond text, it could also be extended to the curation of image or video generation~\cite{almeda2024prompting}, showing thumbnails instead of plain glyphs in the visualization. 

As mentioned in Section~\ref{sec:rw_visual}, GD\&M extends previous work~\cite{suh2023structured} by allowing flexible configuration of arbitrary concepts on continuous dimensions. 
It has a benefit over the previous approach for cases when a single element has a mix of multiple attributes (e.g., a single character is associated with three factions). 
One finding regarding this continuous dimension was that people occasionally disagree with where the tool places elements in the view. This might be because finer granularity in continuous spaces allows users to easily see such disagreements. Our findings indicate that these disagreements facilitate reflection~\cite{ReflectiveCreators} and critical thinking about the user's concepts. 

GD\&M also extends previous work by allowing users to correct AI behaviors directly in the visual spaces. While the interaction itself holds promise, aligning AI behaviors to user-corrected examples was challenging in our version of the tool. 
This might be because there was little room for improvement as the error was already low without any user-added examples.
Future work can explore technical improvements, such as selecting examples that can maximize the performance of recognition and steering~\cite{liu-etal-2022-makes, rubin-etal-2022-learning}.

\subsection{AI-Supported Worldbuilding and Storytelling}

We found that LLMs could support worldbuilding by providing ideas and filling in parts of the world on behalf of the users. However, AI may have both positive and negative effects on creative tasks like worldbuilding. For example, previous work showed that LLM usage could drive users to produce more homogenous responses to a divergent ideation task~\cite{anderson2024homogenization}. 
This could be a problem in a worldbuilding context if users hope to create truly unique worlds. 
As future work, it would be worthwhile to investigate what other problems LLMs might introduce in the context of story writing and what measures might be taken to tackle those problems. 

Study participants expressed a clear desire to use worlds created with \sys{} as the basis for longer-form stories. In particular, participants repeatedly expressed a desire for a specialized timeline view that would enable them to organize world events into a coherent chronology, and in two cases even improvised a timeline view using the existing \sys{} feature set (Figure \ref{fig:patchview_study_timeline}).
AI-supported storytelling might involve many different levels of user control, ranging from humans writing every aspect of the story to the full simulation of the story world by AI~\cite{park2023generative}; previous work hinted at user specification of a high-level story arc~\cite{chung2022talebrush} and participation as a character in the story~\cite{park2023generative}, but many novel interaction paradigms remain for future work to explore. Technically, LLM-based story world simulation would likely face consistency issues~\cite{UnmetNeeds} due to the tendency of LLMs to ``hallucinate''~\cite{ji2023survey}, which would need to be addressed to support coherent storytelling.

\subsection{Technical Alternatives to Prompt Engineering and Closed Models}
\label{sec:discussion_tech}
The prompt engineering techniques we used in \sys{} (including chain-of-thought) yielded good performance on both recognition and steering tasks without any additional training or control techniques. This suggests that current general-purpose LLMs are capable of reasoning effectively about concept relatedness, even along continuous dimensions defined between arbitrary concepts. However, the current \sys{} prototype is limited in its interactivity by relatively high latency. For instance, with two concepts in a view, both steering the generation of a new world element and recognizing its position took longer than 15 seconds. This latency is due partly to the large size of the underlying LLM and partly to our prompting techniques: increasing numbers of world elements, concepts, and user examples cause our prompts to become progressively longer, and chain-of-thought prompting increases the number of tokens generated in response to each prompt, further driving up latency. 
Additionally, current general-purpose LLMs are not specifically tuned for creative applications, resulting in clear weaknesses for creative work~\cite{chakrabarty2024art}. 
These options are often trained on loosely defined preferences~\cite{ouyang2022training}, rather than using rewards more targeted for creative applications.
As they are often closed models, improving on those models would be challenging.

We suggest that future work can explore other options than prompt engineering and closed models. To control generation with the concepts on the continuous dimensions, we can consider the manipulation of task representations in the hidden layers of the LLM~\cite{zou2023representation, vogel2024repeng}. One benefit of this approach is that, once we have the vector about the concept, it does not require tokens for conveying concepts and examples in the prompts. Moreover, if this approach is robust enough, chain-of-thought also might not be necessary. However, future work would need to validate if this approach can effectively steer generation with higher efficiency than prompt engineering. Moreover, how to consider the user-corrected examples with this approach is also moot.
Alternative to closed models would be smaller-sized open models that are fine-tuned to creative use cases, which is becoming more feasible with many open-source LLM options~\cite{jiang2024mixtral, touvron2023llama}. Building upon these open-source LLMs, researchers recently introduced models specialized for creative writing~\cite{wang2024weaver}. Building upon these efforts, avenues we can explore include having a higher quality creative writing dataset with annotations on various aspects of the text quality (e.g., engagingness, novelty, diversity) and tuning models while considering those aspects as losses or rewards~\cite{wu2023fine}.

\subsection{Limitations}
\added{Our tool's usability and functionalities could be improved in the future. For instance, adding filtering functions to the list module would likely help the user find relevant elements when there are many elements to sort through. 
Future versions could also better handle under-specified or irrelevant concepts. The current \sys{} would try to get how the user interprets those concepts from examples provided during the usage. However, such an approach would still have limitations as \sys{} relies on prompt engineering, and alleviating those issues can be future work.}

One limitation of our study is that we could not collect steering results data for rewriting interactions due to technical issues. Moreover, our technical analysis is not the most rigorous but prioritized analyzing data naturally collected during the study. For instance, the user study data might have only covered a subset of genres, settings, and concepts that would frequently used for worldbuilding. \added{The error rates could also have been confounded by the fact that marking errors could incur a small additional cost to users as they need to move elements manually.}
Due to these reasons, future work might involve a more rigorous technical evaluation. This future work could also evaluate other technical implementation options mentioned in Section~\ref{sec:discussion_tech}. 

While \sys{} might be most effective in long-term projects (as it is designed to help users create and organize an expansive fictional world consisting of many distinct elements), our study involved only a single session. 
\added{Future work might investigate the use of \sys{} for long-term projects, including the extension of already existing story worlds.}
Furthermore, we did not compare the design of \sys{} to other alternatives when the user creates their own story world with LLMs; we instead focused more on identifying usage patterns and whether the tool is technically able to achieve its design goals. Future work may investigate \deleted{both long-term usage and }comparison to other tools. 
\section{Conclusion}
We introduce \sys{}, an LLM-powered worldbuilding tool that adopts generative dust and magnet (GD\&M) interactions to support interaction with generative AI.
With GD\&M, \sys{} facilitates sensemaking of generated story elements by placing elements close to concepts of high relevance, similar to how magnets attract iron dust particles.  
It also supports generation steering and AI behavior correction by leveraging the visual space configured by concepts. A user study showed that \sys{} could facilitate understanding the landscape of story world elements and steering of element generation with nuanced intentions that are difficult to express in natural language alone. The interaction of correcting misaligned AI results was intuitive, but those corrections minimally improved the alignment of AI behaviors to the user's perception, indicating one possible direction for future work. We hope \sys{} and GD\&M provide insights on visual interactions for evaluation, specification, and alignment of generative AI behaviors to the user's intention.

\begin{acks}
We thank the many people at Midjourney who provided infrastructural and logistical support for this work. We also thank Yoonjoo Lee, Jordan Huffaker, and Kihoon Son for giving feedback to the early prototype of \sys{}, and user study participants for their valuable insights on the tool.
\end{acks}

\bibliographystyle{ACM-Reference-Format}
\bibliography{main}


\begin{thebibliography}{64}


\ifx \showCODEN    \undefined \def \showCODEN     #1{\unskip}     \fi
\ifx \showDOI      \undefined \def \showDOI       #1{#1}\fi
\ifx \showISBNx    \undefined \def \showISBNx     #1{\unskip}     \fi
\ifx \showISBNxiii \undefined \def \showISBNxiii  #1{\unskip}     \fi
\ifx \showISSN     \undefined \def \showISSN      #1{\unskip}     \fi
\ifx \showLCCN     \undefined \def \showLCCN      #1{\unskip}     \fi
\ifx \shownote     \undefined \def \shownote      #1{#1}          \fi
\ifx \showarticletitle \undefined \def \showarticletitle #1{#1}   \fi
\ifx \showURL      \undefined \def \showURL       {\relax}        \fi
\providecommand\bibfield[2]{#2}
\providecommand\bibinfo[2]{#2}
\providecommand\natexlab[1]{#1}
\providecommand\showeprint[2][]{arXiv:#2}

\bibitem[sud({[n.\,d.]})]%
        {sudowrite_2023}
 \bibinfo{year}{[n.\,d.]}\natexlab{}.
\newblock \bibinfo{title}{Sudowrite}.
\newblock
\newblock
\urldef\tempurl%
\url{https://www.sudowrite.com/}
\showURL{%
\tempurl}


\bibitem[Almeda et~al\mbox{.}(2024)]%
        {almeda2024prompting}
\bibfield{author}{\bibinfo{person}{Shm~Garanganao Almeda}, \bibinfo{person}{J.~D. Zamfirescu-Pereira}, \bibinfo{person}{Kyu~Won Kim}, \bibinfo{person}{Pradeep~Mani Rathnam}, {and} \bibinfo{person}{Bjoern Hartmann}.} \bibinfo{year}{2024}\natexlab{}.
\newblock \bibinfo{title}{Prompting for Discovery: Flexible Sense-Making for AI Art-Making with Dreamsheets}.
\newblock
\newblock
\showeprint[arxiv]{2310.09985}~[cs.HC]


\bibitem[Anderson et~al\mbox{.}(2024)]%
        {anderson2024homogenization}
\bibfield{author}{\bibinfo{person}{Barrett~R. Anderson}, \bibinfo{person}{Jash~Hemant Shah}, {and} \bibinfo{person}{Max Kreminski}.} \bibinfo{year}{2024}\natexlab{}.
\newblock \bibinfo{title}{Homogenization Effects of Large Language Models on Human Creative Ideation}.
\newblock
\newblock
\showeprint[arxiv]{2402.01536}~[cs.HC]


\bibitem[Angert et~al\mbox{.}(2023)]%
        {angert2023spellburst}
\bibfield{author}{\bibinfo{person}{Tyler Angert}, \bibinfo{person}{Miroslav Suzara}, \bibinfo{person}{Jenny Han}, \bibinfo{person}{Christopher Pondoc}, {and} \bibinfo{person}{Hariharan Subramonyam}.} \bibinfo{year}{2023}\natexlab{}.
\newblock \showarticletitle{Spellburst: A Node-Based Interface for Exploratory Creative Coding with Natural Language Prompts}. In \bibinfo{booktitle}{\emph{Proceedings of the 36th Annual ACM Symposium on User Interface Software and Technology}} (San Francisco, CA, USA) \emph{(\bibinfo{series}{UIST '23})}. \bibinfo{publisher}{Association for Computing Machinery}, \bibinfo{address}{New York, NY, USA}, Article \bibinfo{articleno}{100}, \bibinfo{numpages}{22}~pages.
\newblock
\showISBNx{9798400701320}
\urldef\tempurl%
\url{https://doi.org/10.1145/3586183.3606719}
\showDOI{\tempurl}


\bibitem[Anthropic(2023)]%
        {claude2}
\bibfield{author}{\bibinfo{person}{Anthropic}.} \bibinfo{year}{2023}\natexlab{}.
\newblock \bibinfo{title}{Model Card and Evaluations for Claude Models}.
\newblock
\newblock
\urldef\tempurl%
\url{https://www-files.anthropic.com/production/images/Model-Card-Claude-2.pdf}
\showURL{%
\tempurl}


\bibitem[Arawjo et~al\mbox{.}(2023)]%
        {arawjo2023chainforge}
\bibfield{author}{\bibinfo{person}{Ian Arawjo}, \bibinfo{person}{Chelse Swoopes}, \bibinfo{person}{Priyan Vaithilingam}, \bibinfo{person}{Martin Wattenberg}, {and} \bibinfo{person}{Elena Glassman}.} \bibinfo{year}{2023}\natexlab{}.
\newblock \bibinfo{title}{ChainForge: A Visual Toolkit for Prompt Engineering and LLM Hypothesis Testing}.
\newblock
\newblock
\showeprint[arxiv]{2309.09128}~[cs.HC]


\bibitem[Biermann et~al\mbox{.}(2022)]%
        {biermann2022from}
\bibfield{author}{\bibinfo{person}{Oloff~C. Biermann}, \bibinfo{person}{Ning~F. Ma}, {and} \bibinfo{person}{Dongwook Yoon}.} \bibinfo{year}{2022}\natexlab{}.
\newblock \showarticletitle{From Tool to Companion: Storywriters Want AI Writers to Respect Their Personal Values and Writing Strategies}. In \bibinfo{booktitle}{\emph{Proceedings of the 2022 ACM Designing Interactive Systems Conference}} (Virtual Event, Australia) \emph{(\bibinfo{series}{DIS '22})}. \bibinfo{publisher}{Association for Computing Machinery}, \bibinfo{address}{New York, NY, USA}, \bibinfo{pages}{1209–1227}.
\newblock
\showISBNx{9781450393584}
\urldef\tempurl%
\url{https://doi.org/10.1145/3532106.3533506}
\showDOI{\tempurl}


\bibitem[Brown et~al\mbox{.}(2020)]%
        {brown2020language}
\bibfield{author}{\bibinfo{person}{Tom Brown}, \bibinfo{person}{Benjamin Mann}, \bibinfo{person}{Nick Ryder}, \bibinfo{person}{Melanie Subbiah}, \bibinfo{person}{Jared~D Kaplan}, \bibinfo{person}{Prafulla Dhariwal}, \bibinfo{person}{Arvind Neelakantan}, \bibinfo{person}{Pranav Shyam}, \bibinfo{person}{Girish Sastry}, \bibinfo{person}{Amanda Askell}, \bibinfo{person}{Sandhini Agarwal}, \bibinfo{person}{Ariel Herbert-Voss}, \bibinfo{person}{Gretchen Krueger}, \bibinfo{person}{Tom Henighan}, \bibinfo{person}{Rewon Child}, \bibinfo{person}{Aditya Ramesh}, \bibinfo{person}{Daniel Ziegler}, \bibinfo{person}{Jeffrey Wu}, \bibinfo{person}{Clemens Winter}, \bibinfo{person}{Chris Hesse}, \bibinfo{person}{Mark Chen}, \bibinfo{person}{Eric Sigler}, \bibinfo{person}{Mateusz Litwin}, \bibinfo{person}{Scott Gray}, \bibinfo{person}{Benjamin Chess}, \bibinfo{person}{Jack Clark}, \bibinfo{person}{Christopher Berner}, \bibinfo{person}{Sam McCandlish}, \bibinfo{person}{Alec Radford}, \bibinfo{person}{Ilya Sutskever}, {and}
  \bibinfo{person}{Dario Amodei}.} \bibinfo{year}{2020}\natexlab{}.
\newblock \showarticletitle{Language Models are Few-Shot Learners}. In \bibinfo{booktitle}{\emph{Advances in Neural Information Processing Systems}}, \bibfield{editor}{\bibinfo{person}{H.~Larochelle}, \bibinfo{person}{M.~Ranzato}, \bibinfo{person}{R.~Hadsell}, \bibinfo{person}{M.~F. Balcan}, {and} \bibinfo{person}{H.~Lin}} (Eds.), Vol.~\bibinfo{volume}{33}. \bibinfo{publisher}{Curran Associates, Inc.}, \bibinfo{pages}{1877--1901}.
\newblock
\urldef\tempurl%
\url{https://proceedings.neurips.cc/paper/2020/file/1457c0d6bfcb4967418bfb8ac142f64a-Paper.pdf}
\showURL{%
\tempurl}


\bibitem[Calderwood et~al\mbox{.}(2020)]%
        {calderwood2020hownovelists}
\bibfield{author}{\bibinfo{person}{Alex Calderwood}, \bibinfo{person}{Vivian Qiu}, \bibinfo{person}{K. Gero}, {and} \bibinfo{person}{Lydia~B. Chilton}.} \bibinfo{year}{2020}\natexlab{}.
\newblock \showarticletitle{How Novelists Use Generative Language Models: An Exploratory User Study}. In \bibinfo{booktitle}{\emph{HAI-GEN+user2agent@IUI}}.
\newblock
\urldef\tempurl%
\url{https://api.semanticscholar.org/CorpusID:233479959}
\showURL{%
\tempurl}


\bibitem[Chakrabarty et~al\mbox{.}(2024)]%
        {chakrabarty2024art}
\bibfield{author}{\bibinfo{person}{Tuhin Chakrabarty}, \bibinfo{person}{Philippe Laban}, \bibinfo{person}{Divyansh Agarwal}, \bibinfo{person}{Smaranda Muresan}, {and} \bibinfo{person}{Chien-Sheng Wu}.} \bibinfo{year}{2024}\natexlab{}.
\newblock \bibinfo{title}{Art or Artifice? Large Language Models and the False Promise of Creativity}.
\newblock
\newblock
\showeprint[arxiv]{2309.14556}~[cs.CL]


\bibitem[Chen et~al\mbox{.}(2018)]%
        {chen2018anchorviz}
\bibfield{author}{\bibinfo{person}{Nan-Chen Chen}, \bibinfo{person}{Jina Suh}, \bibinfo{person}{Johan Verwey}, \bibinfo{person}{Gonzalo Ramos}, \bibinfo{person}{Steven Drucker}, {and} \bibinfo{person}{Patrice Simard}.} \bibinfo{year}{2018}\natexlab{}.
\newblock \showarticletitle{AnchorViz: Facilitating Classifier Error Discovery through Interactive Semantic Data Exploration}. In \bibinfo{booktitle}{\emph{23rd International Conference on Intelligent User Interfaces}} (Tokyo, Japan) \emph{(\bibinfo{series}{IUI '18})}. \bibinfo{publisher}{Association for Computing Machinery}, \bibinfo{address}{New York, NY, USA}, \bibinfo{pages}{269–280}.
\newblock
\showISBNx{9781450349451}
\urldef\tempurl%
\url{https://doi.org/10.1145/3172944.3172950}
\showDOI{\tempurl}


\bibitem[Cherry and Latulipe(2014)]%
        {cherry2014quantifying}
\bibfield{author}{\bibinfo{person}{Erin Cherry} {and} \bibinfo{person}{Celine Latulipe}.} \bibinfo{year}{2014}\natexlab{}.
\newblock \showarticletitle{Quantifying the Creativity Support of Digital Tools through the Creativity Support Index}.
\newblock \bibinfo{journal}{\emph{ACM Trans. Comput.-Hum. Interact.}} \bibinfo{volume}{21}, \bibinfo{number}{4}, Article \bibinfo{articleno}{21} (\bibinfo{date}{jun} \bibinfo{year}{2014}), \bibinfo{numpages}{25}~pages.
\newblock
\showISSN{1073-0516}
\urldef\tempurl%
\url{https://doi.org/10.1145/2617588}
\showDOI{\tempurl}


\bibitem[Chou et~al\mbox{.}(2023)]%
        {chou2023talestream}
\bibfield{author}{\bibinfo{person}{Jean-Pe\"{\i}c Chou}, \bibinfo{person}{Alexa~Fay Siu}, \bibinfo{person}{Nedim Lipka}, \bibinfo{person}{Ryan Rossi}, \bibinfo{person}{Franck Dernoncourt}, {and} \bibinfo{person}{Maneesh Agrawala}.} \bibinfo{year}{2023}\natexlab{}.
\newblock \showarticletitle{TaleStream: Supporting Story Ideation with Trope Knowledge}. In \bibinfo{booktitle}{\emph{Proceedings of the 36th Annual ACM Symposium on User Interface Software and Technology}} (San Francisco, CA, USA) \emph{(\bibinfo{series}{UIST '23})}. \bibinfo{publisher}{Association for Computing Machinery}, \bibinfo{address}{New York, NY, USA}, Article \bibinfo{articleno}{52}, \bibinfo{numpages}{12}~pages.
\newblock
\showISBNx{9798400701320}
\urldef\tempurl%
\url{https://doi.org/10.1145/3586183.3606807}
\showDOI{\tempurl}


\bibitem[Chowdhery et~al\mbox{.}(2022)]%
        {chowdhery2022palm}
\bibfield{author}{\bibinfo{person}{Aakanksha Chowdhery}, \bibinfo{person}{Sharan Narang}, \bibinfo{person}{Jacob Devlin}, \bibinfo{person}{Maarten Bosma}, \bibinfo{person}{Gaurav Mishra}, \bibinfo{person}{Adam Roberts}, \bibinfo{person}{Paul Barham}, \bibinfo{person}{Hyung~Won Chung}, \bibinfo{person}{Charles Sutton}, \bibinfo{person}{Sebastian Gehrmann}, \bibinfo{person}{Parker Schuh}, \bibinfo{person}{Kensen Shi}, \bibinfo{person}{Sasha Tsvyashchenko}, \bibinfo{person}{Joshua Maynez}, \bibinfo{person}{Abhishek Rao}, \bibinfo{person}{Parker Barnes}, \bibinfo{person}{Yi Tay}, \bibinfo{person}{Noam Shazeer}, \bibinfo{person}{Vinodkumar Prabhakaran}, \bibinfo{person}{Emily Reif}, \bibinfo{person}{Nan Du}, \bibinfo{person}{Ben Hutchinson}, \bibinfo{person}{Reiner Pope}, \bibinfo{person}{James Bradbury}, \bibinfo{person}{Jacob Austin}, \bibinfo{person}{Michael Isard}, \bibinfo{person}{Guy Gur-Ari}, \bibinfo{person}{Pengcheng Yin}, \bibinfo{person}{Toju Duke}, \bibinfo{person}{Anselm Levskaya},
  \bibinfo{person}{Sanjay Ghemawat}, \bibinfo{person}{Sunipa Dev}, \bibinfo{person}{Henryk Michalewski}, \bibinfo{person}{Xavier Garcia}, \bibinfo{person}{Vedant Misra}, \bibinfo{person}{Kevin Robinson}, \bibinfo{person}{Liam Fedus}, \bibinfo{person}{Denny Zhou}, \bibinfo{person}{Daphne Ippolito}, \bibinfo{person}{David Luan}, \bibinfo{person}{Hyeontaek Lim}, \bibinfo{person}{Barret Zoph}, \bibinfo{person}{Alexander Spiridonov}, \bibinfo{person}{Ryan Sepassi}, \bibinfo{person}{David Dohan}, \bibinfo{person}{Shivani Agrawal}, \bibinfo{person}{Mark Omernick}, \bibinfo{person}{Andrew~M. Dai}, \bibinfo{person}{Thanumalayan~Sankaranarayana Pillai}, \bibinfo{person}{Marie Pellat}, \bibinfo{person}{Aitor Lewkowycz}, \bibinfo{person}{Erica Moreira}, \bibinfo{person}{Rewon Child}, \bibinfo{person}{Oleksandr Polozov}, \bibinfo{person}{Katherine Lee}, \bibinfo{person}{Zongwei Zhou}, \bibinfo{person}{Xuezhi Wang}, \bibinfo{person}{Brennan Saeta}, \bibinfo{person}{Mark Diaz}, \bibinfo{person}{Orhan Firat},
  \bibinfo{person}{Michele Catasta}, \bibinfo{person}{Jason Wei}, \bibinfo{person}{Kathy Meier-Hellstern}, \bibinfo{person}{Douglas Eck}, \bibinfo{person}{Jeff Dean}, \bibinfo{person}{Slav Petrov}, {and} \bibinfo{person}{Noah Fiedel}.} \bibinfo{year}{2022}\natexlab{}.
\newblock \bibinfo{title}{PaLM: Scaling Language Modeling with Pathways}.
\newblock
\newblock
\showeprint[arxiv]{2204.02311}~[cs.CL]


\bibitem[Chung and Adar(2023)]%
        {chung2023promptpaint}
\bibfield{author}{\bibinfo{person}{John Joon~Young Chung} {and} \bibinfo{person}{Eytan Adar}.} \bibinfo{year}{2023}\natexlab{}.
\newblock \showarticletitle{PromptPaint: Steering Text-to-Image Generation Through Paint Medium-like Interactions}. In \bibinfo{booktitle}{\emph{Proceedings of the 36th Annual ACM Symposium on User Interface Software and Technology}} (San Francisco, CA, USA) \emph{(\bibinfo{series}{UIST '23})}. \bibinfo{publisher}{Association for Computing Machinery}, \bibinfo{address}{New York, NY, USA}, Article \bibinfo{articleno}{6}, \bibinfo{numpages}{17}~pages.
\newblock
\showISBNx{9798400701320}
\urldef\tempurl%
\url{https://doi.org/10.1145/3586183.3606777}
\showDOI{\tempurl}


\bibitem[Chung et~al\mbox{.}(2022a)]%
        {chung2022artist}
\bibfield{author}{\bibinfo{person}{John Joon~Young Chung}, \bibinfo{person}{Shiqing He}, {and} \bibinfo{person}{Eytan Adar}.} \bibinfo{year}{2022}\natexlab{a}.
\newblock \showarticletitle{Artist Support Networks: Implications for Future Creativity Support Tools}. In \bibinfo{booktitle}{\emph{Proceedings of the 2022 ACM Designing Interactive Systems Conference}} (Virtual Event, Australia) \emph{(\bibinfo{series}{DIS '22})}. \bibinfo{publisher}{Association for Computing Machinery}, \bibinfo{address}{New York, NY, USA}, \bibinfo{pages}{232–246}.
\newblock
\showISBNx{9781450393584}
\urldef\tempurl%
\url{https://doi.org/10.1145/3532106.3533505}
\showDOI{\tempurl}


\bibitem[Chung et~al\mbox{.}(2022b)]%
        {chung2022talebrush}
\bibfield{author}{\bibinfo{person}{John Joon~Young Chung}, \bibinfo{person}{Wooseok Kim}, \bibinfo{person}{Kang~Min Yoo}, \bibinfo{person}{Hwaran Lee}, \bibinfo{person}{Eytan Adar}, {and} \bibinfo{person}{Minsuk Chang}.} \bibinfo{year}{2022}\natexlab{b}.
\newblock \showarticletitle{TaleBrush: Sketching Stories with Generative Pretrained Language Models}. In \bibinfo{booktitle}{\emph{Proceedings of the 2022 CHI Conference on Human Factors in Computing Systems}} (New Orleans, LA, USA) \emph{(\bibinfo{series}{CHI '22})}. \bibinfo{publisher}{Association for Computing Machinery}, \bibinfo{address}{New York, NY, USA}, Article \bibinfo{articleno}{209}, \bibinfo{numpages}{19}~pages.
\newblock
\showISBNx{9781450391573}
\urldef\tempurl%
\url{https://doi.org/10.1145/3491102.3501819}
\showDOI{\tempurl}


\bibitem[Chung et~al\mbox{.}(2021)]%
        {chung2021beyond}
\bibfield{author}{\bibinfo{person}{John Joon~Young Chung}, \bibinfo{person}{Hijung~Valentina Shin}, \bibinfo{person}{Haijun Xia}, \bibinfo{person}{Li-yi Wei}, {and} \bibinfo{person}{Rubaiat~Habib Kazi}.} \bibinfo{year}{2021}\natexlab{}.
\newblock \showarticletitle{Beyond Show of Hands: Engaging Viewers via Expressive and Scalable Visual Communication in Live Streaming}. In \bibinfo{booktitle}{\emph{Proceedings of the 2021 CHI Conference on Human Factors in Computing Systems}} (Yokohama, Japan) \emph{(\bibinfo{series}{CHI '21})}. \bibinfo{publisher}{Association for Computing Machinery}, \bibinfo{address}{New York, NY, USA}, Article \bibinfo{articleno}{109}, \bibinfo{numpages}{14}~pages.
\newblock
\showISBNx{9781450380966}
\urldef\tempurl%
\url{https://doi.org/10.1145/3411764.3445419}
\showDOI{\tempurl}


\bibitem[Clark et~al\mbox{.}(2018)]%
        {clark2028creative}
\bibfield{author}{\bibinfo{person}{Elizabeth Clark}, \bibinfo{person}{Anne~Spencer Ross}, \bibinfo{person}{Chenhao Tan}, \bibinfo{person}{Yangfeng Ji}, {and} \bibinfo{person}{Noah~A. Smith}.} \bibinfo{year}{2018}\natexlab{}.
\newblock \showarticletitle{Creative Writing with a Machine in the Loop: Case Studies on Slogans and Stories}. In \bibinfo{booktitle}{\emph{23rd International Conference on Intelligent User Interfaces}} (Tokyo, Japan) \emph{(\bibinfo{series}{IUI '18})}. \bibinfo{publisher}{Association for Computing Machinery}, \bibinfo{address}{New York, NY, USA}, \bibinfo{pages}{329–340}.
\newblock
\showISBNx{9781450349451}
\urldef\tempurl%
\url{https://doi.org/10.1145/3172944.3172983}
\showDOI{\tempurl}


\bibitem[Dang et~al\mbox{.}(2023)]%
        {dang2023worldsmith}
\bibfield{author}{\bibinfo{person}{Hai Dang}, \bibinfo{person}{Frederik Brudy}, \bibinfo{person}{George Fitzmaurice}, {and} \bibinfo{person}{Fraser Anderson}.} \bibinfo{year}{2023}\natexlab{}.
\newblock \showarticletitle{WorldSmith: Iterative and Expressive Prompting for World Building with a Generative AI}. In \bibinfo{booktitle}{\emph{Proceedings of the 36th Annual ACM Symposium on User Interface Software and Technology}} (San Francisco, CA, USA) \emph{(\bibinfo{series}{UIST '23})}. \bibinfo{publisher}{Association for Computing Machinery}, \bibinfo{address}{New York, NY, USA}, Article \bibinfo{articleno}{63}, \bibinfo{numpages}{17}~pages.
\newblock
\showISBNx{9798400701320}
\urldef\tempurl%
\url{https://doi.org/10.1145/3586183.3606772}
\showDOI{\tempurl}


\bibitem[Fast and Örnebring(2017)]%
        {fast2017transmedia}
\bibfield{author}{\bibinfo{person}{Karin Fast} {and} \bibinfo{person}{Henrik Örnebring}.} \bibinfo{year}{2017}\natexlab{}.
\newblock \showarticletitle{Transmedia world-building: The Shadow (1931–present) and Transformers (1984–present)}.
\newblock \bibinfo{journal}{\emph{International Journal of Cultural Studies}} \bibinfo{volume}{20}, \bibinfo{number}{6} (\bibinfo{year}{2017}), \bibinfo{pages}{636--652}.
\newblock
\urldef\tempurl%
\url{https://doi.org/10.1177/1367877915605887}
\showDOI{\tempurl}
\showeprint{https://doi.org/10.1177/1367877915605887}


\bibitem[Gero et~al\mbox{.}(2023)]%
        {gero2023social}
\bibfield{author}{\bibinfo{person}{Katy~Ilonka Gero}, \bibinfo{person}{Tao Long}, {and} \bibinfo{person}{Lydia~B Chilton}.} \bibinfo{year}{2023}\natexlab{}.
\newblock \showarticletitle{Social Dynamics of AI Support in Creative Writing}. In \bibinfo{booktitle}{\emph{Proceedings of the 2023 CHI Conference on Human Factors in Computing Systems}} (Hamburg, Germany) \emph{(\bibinfo{series}{CHI '23})}. \bibinfo{publisher}{Association for Computing Machinery}, \bibinfo{address}{New York, NY, USA}, Article \bibinfo{articleno}{245}, \bibinfo{numpages}{15}~pages.
\newblock
\showISBNx{9781450394215}
\urldef\tempurl%
\url{https://doi.org/10.1145/3544548.3580782}
\showDOI{\tempurl}


\bibitem[Gong et~al\mbox{.}(2023)]%
        {gong2023interactive}
\bibfield{author}{\bibinfo{person}{Yuan Gong}, \bibinfo{person}{Youxin Pang}, \bibinfo{person}{Xiaodong Cun}, \bibinfo{person}{Menghan Xia}, \bibinfo{person}{Yingqing He}, \bibinfo{person}{Haoxin Chen}, \bibinfo{person}{Longyue Wang}, \bibinfo{person}{Yong Zhang}, \bibinfo{person}{Xintao Wang}, \bibinfo{person}{Ying Shan}, {and} \bibinfo{person}{Yujiu Yang}.} \bibinfo{year}{2023}\natexlab{}.
\newblock \showarticletitle{Interactive Story Visualization with Multiple Characters}. In \bibinfo{booktitle}{\emph{SIGGRAPH Asia 2023 Conference Papers}} (Sydney, NSW, Australia) \emph{(\bibinfo{series}{SA '23})}. \bibinfo{publisher}{Association for Computing Machinery}, \bibinfo{address}{New York, NY, USA}, Article \bibinfo{articleno}{101}, \bibinfo{numpages}{10}~pages.
\newblock
\showISBNx{9798400703157}
\urldef\tempurl%
\url{https://doi.org/10.1145/3610548.3618184}
\showDOI{\tempurl}


\bibitem[Hergenrader(2018)]%
        {hergenrader2018collaborative}
\bibfield{author}{\bibinfo{person}{T. Hergenrader}.} \bibinfo{year}{2018}\natexlab{}.
\newblock \bibinfo{booktitle}{\emph{Collaborative Worldbuilding for Writers and Gamers}}.
\newblock \bibinfo{publisher}{Bloomsbury Academic}.
\newblock
\showISBNx{9781350016668}
\showLCCN{2018040236}
\urldef\tempurl%
\url{https://books.google.co.kr/books?id=z-_7swEACAAJ}
\showURL{%
\tempurl}


\bibitem[Hoque et~al\mbox{.}(2023)]%
        {hoque2023portrayal}
\bibfield{author}{\bibinfo{person}{Md~Naimul Hoque}, \bibinfo{person}{Bhavya Ghai}, \bibinfo{person}{Kari Kraus}, {and} \bibinfo{person}{Niklas Elmqvist}.} \bibinfo{year}{2023}\natexlab{}.
\newblock \showarticletitle{Portrayal: Leveraging NLP and Visualization for Analyzing Fictional Characters}. In \bibinfo{booktitle}{\emph{Proceedings of the 2023 ACM Designing Interactive Systems Conference}} (Pittsburgh, PA, USA) \emph{(\bibinfo{series}{DIS '23})}. \bibinfo{publisher}{Association for Computing Machinery}, \bibinfo{address}{New York, NY, USA}, \bibinfo{pages}{74–94}.
\newblock
\showISBNx{9781450398930}
\urldef\tempurl%
\url{https://doi.org/10.1145/3563657.3596000}
\showDOI{\tempurl}


\bibitem[Ji et~al\mbox{.}(2023)]%
        {ji2023survey}
\bibfield{author}{\bibinfo{person}{Ziwei Ji}, \bibinfo{person}{Nayeon Lee}, \bibinfo{person}{Rita Frieske}, \bibinfo{person}{Tiezheng Yu}, \bibinfo{person}{Dan Su}, \bibinfo{person}{Yan Xu}, \bibinfo{person}{Etsuko Ishii}, \bibinfo{person}{Ye~Jin Bang}, \bibinfo{person}{Andrea Madotto}, {and} \bibinfo{person}{Pascale Fung}.} \bibinfo{year}{2023}\natexlab{}.
\newblock \showarticletitle{Survey of Hallucination in Natural Language Generation}.
\newblock \bibinfo{journal}{\emph{ACM Comput. Surv.}} \bibinfo{volume}{55}, \bibinfo{number}{12}, Article \bibinfo{articleno}{248} (\bibinfo{date}{mar} \bibinfo{year}{2023}), \bibinfo{numpages}{38}~pages.
\newblock
\showISSN{0360-0300}
\urldef\tempurl%
\url{https://doi.org/10.1145/3571730}
\showDOI{\tempurl}


\bibitem[Jiang et~al\mbox{.}(2024)]%
        {jiang2024mixtral}
\bibfield{author}{\bibinfo{person}{Albert~Q. Jiang}, \bibinfo{person}{Alexandre Sablayrolles}, \bibinfo{person}{Antoine Roux}, \bibinfo{person}{Arthur Mensch}, \bibinfo{person}{Blanche Savary}, \bibinfo{person}{Chris Bamford}, \bibinfo{person}{Devendra~Singh Chaplot}, \bibinfo{person}{Diego de~las Casas}, \bibinfo{person}{Emma~Bou Hanna}, \bibinfo{person}{Florian Bressand}, \bibinfo{person}{Gianna Lengyel}, \bibinfo{person}{Guillaume Bour}, \bibinfo{person}{Guillaume Lample}, \bibinfo{person}{Lélio~Renard Lavaud}, \bibinfo{person}{Lucile Saulnier}, \bibinfo{person}{Marie-Anne Lachaux}, \bibinfo{person}{Pierre Stock}, \bibinfo{person}{Sandeep Subramanian}, \bibinfo{person}{Sophia Yang}, \bibinfo{person}{Szymon Antoniak}, \bibinfo{person}{Teven~Le Scao}, \bibinfo{person}{Théophile Gervet}, \bibinfo{person}{Thibaut Lavril}, \bibinfo{person}{Thomas Wang}, \bibinfo{person}{Timothée Lacroix}, {and} \bibinfo{person}{William~El Sayed}.} \bibinfo{year}{2024}\natexlab{}.
\newblock \bibinfo{title}{Mixtral of Experts}.
\newblock
\newblock
\showeprint[arxiv]{2401.04088}~[cs.LG]


\bibitem[Jiang et~al\mbox{.}(2023)]%
        {jiang2023graphologue}
\bibfield{author}{\bibinfo{person}{Peiling Jiang}, \bibinfo{person}{Jude Rayan}, \bibinfo{person}{Steven~P. Dow}, {and} \bibinfo{person}{Haijun Xia}.} \bibinfo{year}{2023}\natexlab{}.
\newblock \showarticletitle{Graphologue: Exploring Large Language Model Responses with Interactive Diagrams}. In \bibinfo{booktitle}{\emph{Proceedings of the 36th Annual ACM Symposium on User Interface Software and Technology}} (San Francisco, CA, USA) \emph{(\bibinfo{series}{UIST '23})}. \bibinfo{publisher}{Association for Computing Machinery}, \bibinfo{address}{New York, NY, USA}, Article \bibinfo{articleno}{3}, \bibinfo{numpages}{20}~pages.
\newblock
\showISBNx{9798400701320}
\urldef\tempurl%
\url{https://doi.org/10.1145/3586183.3606737}
\showDOI{\tempurl}


\bibitem[Kim et~al\mbox{.}(2024)]%
        {kim2024authors}
\bibfield{author}{\bibinfo{person}{Taewook Kim}, \bibinfo{person}{Hyomin Han}, \bibinfo{person}{Eytan Adar}, \bibinfo{person}{Matthew Kay}, {and} \bibinfo{person}{John Joon~Young Chung}.} \bibinfo{year}{2024}\natexlab{}.
\newblock \showarticletitle{Authors' Values and Attitudes Towards AI-bridged Scalable Personalization of Creative Language Arts}. In \bibinfo{booktitle}{\emph{Proceedings of the 2024 CHI Conference on Human Factors in Computing Systems}} (Honolulu, HI, USA) \emph{(\bibinfo{series}{CHI '24})}. \bibinfo{publisher}{Association for Computing Machinery}, \bibinfo{address}{New York, NY, USA}.
\newblock


\bibitem[Kim et~al\mbox{.}(2023)]%
        {kim2023cells}
\bibfield{author}{\bibinfo{person}{Tae~Soo Kim}, \bibinfo{person}{Yoonjoo Lee}, \bibinfo{person}{Minsuk Chang}, {and} \bibinfo{person}{Juho Kim}.} \bibinfo{year}{2023}\natexlab{}.
\newblock \showarticletitle{Cells, Generators, and Lenses: Design Framework for Object-Oriented Interaction with Large Language Models}. In \bibinfo{booktitle}{\emph{Proceedings of the 36th Annual ACM Symposium on User Interface Software and Technology}} (San Francisco, CA, USA) \emph{(\bibinfo{series}{UIST '23})}. \bibinfo{publisher}{Association for Computing Machinery}, \bibinfo{address}{New York, NY, USA}, Article \bibinfo{articleno}{4}, \bibinfo{numpages}{18}~pages.
\newblock
\showISBNx{9798400701320}
\urldef\tempurl%
\url{https://doi.org/10.1145/3586183.3606833}
\showDOI{\tempurl}


\bibitem[Kreminski et~al\mbox{.}(2022)]%
        {kreminski2022loose}
\bibfield{author}{\bibinfo{person}{Max Kreminski}, \bibinfo{person}{Melanie Dickinson}, \bibinfo{person}{Noah Wardrip-Fruin}, {and} \bibinfo{person}{Michael Mateas}.} \bibinfo{year}{2022}\natexlab{}.
\newblock \showarticletitle{Loose Ends: A Mixed-Initiative Creative Interface for Playful Storytelling}.
\newblock \bibinfo{journal}{\emph{Proceedings of the AAAI Conference on Artificial Intelligence and Interactive Digital Entertainment}} \bibinfo{volume}{18}, \bibinfo{number}{1} (\bibinfo{date}{Oct.} \bibinfo{year}{2022}), \bibinfo{pages}{120--128}.
\newblock
\urldef\tempurl%
\url{https://doi.org/10.1609/aiide.v18i1.21955}
\showDOI{\tempurl}


\bibitem[Kreminski and Martens(2022)]%
        {UnmetNeeds}
\bibfield{author}{\bibinfo{person}{Max Kreminski} {and} \bibinfo{person}{Chris Martens}.} \bibinfo{year}{2022}\natexlab{}.
\newblock \showarticletitle{Unmet creativity support needs in computationally supported creative writing}. In \bibinfo{booktitle}{\emph{Proceedings of the First Workshop on Intelligent and Interactive Writing Assistants (In2Writing 2022)}}. \bibinfo{pages}{74--82}.
\newblock


\bibitem[Kreminski and Mateas(2021)]%
        {ReflectiveCreators}
\bibfield{author}{\bibinfo{person}{Max Kreminski} {and} \bibinfo{person}{Michael Mateas}.} \bibinfo{year}{2021}\natexlab{}.
\newblock \showarticletitle{Reflective Creators}. In \bibinfo{booktitle}{\emph{International Conference on Computational Creativity}}. \bibinfo{pages}{309--318}.
\newblock


\bibitem[Lee et~al\mbox{.}(2024)]%
        {lee2024dsiiwa}
\bibfield{author}{\bibinfo{person}{Mina Lee}, \bibinfo{person}{Katy~Ilonka Gero}, \bibinfo{person}{John Joon~Young Chung}, \bibinfo{person}{Simon~Buckingham Shum}, \bibinfo{person}{Vipul Raheja}, \bibinfo{person}{Hua Shen}, \bibinfo{person}{Subhashini Venugopalan}, \bibinfo{person}{Thiemo Wambsganss}, \bibinfo{person}{David Zhou}, \bibinfo{person}{Emad~A. Alghamdi}, \bibinfo{person}{Tal August}, \bibinfo{person}{Avinash Bhat}, \bibinfo{person}{Madiha~Zahrah Choksi}, \bibinfo{person}{Senjuti Dutta}, \bibinfo{person}{Jin~L.C. Guo}, \bibinfo{person}{Md~Naimul Hoque}, \bibinfo{person}{Yewon Kim}, \bibinfo{person}{Simon Knight}, \bibinfo{person}{Seyed~Parsa Neshaei}, \bibinfo{person}{Antonette Shibani}, \bibinfo{person}{Disha Shrivastava}, \bibinfo{person}{Lila Shroff}, \bibinfo{person}{Agnia Sergeyuk}, \bibinfo{person}{Jessi Stark}, \bibinfo{person}{Sarah Sterman}, \bibinfo{person}{Sitong Wang}, \bibinfo{person}{Antoine Bosselut}, \bibinfo{person}{Daniel Buschek}, \bibinfo{person}{Joseph~Chee Chang},
  \bibinfo{person}{Sherol Chen}, \bibinfo{person}{Max Kreminski}, \bibinfo{person}{Joonsuk Park}, \bibinfo{person}{Roy Pea}, \bibinfo{person}{Eugenia Ha~Rim Rho}, \bibinfo{person}{Zejiang Shen}, {and} \bibinfo{person}{Pao Siangliulue}.} \bibinfo{year}{2024}\natexlab{}.
\newblock \showarticletitle{A Design Space for Intelligent and Interactive Writing Assistants}. In \bibinfo{booktitle}{\emph{Proceedings of the 2024 CHI Conference on Human Factors in Computing Systems}} (Honolulu, HI, USA) \emph{(\bibinfo{series}{CHI '24})}. \bibinfo{publisher}{Association for Computing Machinery}, \bibinfo{address}{New York, NY, USA}.
\newblock


\bibitem[Lee et~al\mbox{.}(2022)]%
        {lee2022coauthor}
\bibfield{author}{\bibinfo{person}{Mina Lee}, \bibinfo{person}{Percy Liang}, {and} \bibinfo{person}{Qian Yang}.} \bibinfo{year}{2022}\natexlab{}.
\newblock \showarticletitle{CoAuthor: Designing a Human-AI Collaborative Writing Dataset for Exploring Language Model Capabilities}. In \bibinfo{booktitle}{\emph{Proceedings of the 2022 CHI Conference on Human Factors in Computing Systems}} (New Orleans, LA, USA) \emph{(\bibinfo{series}{CHI '22})}. \bibinfo{publisher}{Association for Computing Machinery}, \bibinfo{address}{New York, NY, USA}, Article \bibinfo{articleno}{388}, \bibinfo{numpages}{19}~pages.
\newblock
\showISBNx{9781450391573}
\urldef\tempurl%
\url{https://doi.org/10.1145/3491102.3502030}
\showDOI{\tempurl}


\bibitem[Liu et~al\mbox{.}(2022)]%
        {liu-etal-2022-makes}
\bibfield{author}{\bibinfo{person}{Jiachang Liu}, \bibinfo{person}{Dinghan Shen}, \bibinfo{person}{Yizhe Zhang}, \bibinfo{person}{Bill Dolan}, \bibinfo{person}{Lawrence Carin}, {and} \bibinfo{person}{Weizhu Chen}.} \bibinfo{year}{2022}\natexlab{}.
\newblock \showarticletitle{What Makes Good In-Context Examples for {GPT}-3?}. In \bibinfo{booktitle}{\emph{Proceedings of Deep Learning Inside Out (DeeLIO 2022): The 3rd Workshop on Knowledge Extraction and Integration for Deep Learning Architectures}}, \bibfield{editor}{\bibinfo{person}{Eneko Agirre}, \bibinfo{person}{Marianna Apidianaki}, {and} \bibinfo{person}{Ivan Vuli{\'c}}} (Eds.). \bibinfo{publisher}{Association for Computational Linguistics}, \bibinfo{address}{Dublin, Ireland and Online}, \bibinfo{pages}{100--114}.
\newblock
\urldef\tempurl%
\url{https://doi.org/10.18653/v1/2022.deelio-1.10}
\showDOI{\tempurl}


\bibitem[Livingstone(1986)]%
        {livingstone1986dicing}
\bibfield{author}{\bibinfo{person}{I. Livingstone}.} \bibinfo{year}{1986}\natexlab{}.
\newblock \bibinfo{booktitle}{\emph{Dicing with Dragons: An Introduction to Role-Playing Games}}.
\newblock \bibinfo{publisher}{Penguin Group USA, Incorporated}.
\newblock
\showISBNx{9780451144898}
\urldef\tempurl%
\url{https://books.google.co.kr/books?id=uBqXPwAACAAJ}
\showURL{%
\tempurl}


\bibitem[Louie et~al\mbox{.}(2020)]%
        {louie2020novice}
\bibfield{author}{\bibinfo{person}{Ryan Louie}, \bibinfo{person}{Andy Coenen}, \bibinfo{person}{Cheng~Zhi Huang}, \bibinfo{person}{Michael Terry}, {and} \bibinfo{person}{Carrie~J. Cai}.} \bibinfo{year}{2020}\natexlab{}.
\newblock \showarticletitle{Novice-AI Music Co-Creation via AI-Steering Tools for Deep Generative Models}. In \bibinfo{booktitle}{\emph{Proceedings of the 2020 CHI Conference on Human Factors in Computing Systems}} (Honolulu, HI, USA) \emph{(\bibinfo{series}{CHI '20})}. \bibinfo{publisher}{Association for Computing Machinery}, \bibinfo{address}{New York, NY, USA}, \bibinfo{pages}{1–13}.
\newblock
\showISBNx{9781450367080}
\urldef\tempurl%
\url{https://doi.org/10.1145/3313831.3376739}
\showDOI{\tempurl}


\bibitem[Maj(2015)]%
        {maj2015transmedial}
\bibfield{author}{\bibinfo{person}{Krzysztof~M Maj}.} \bibinfo{year}{2015}\natexlab{}.
\newblock \showarticletitle{Transmedial world-building in fictional narratives}.
\newblock \bibinfo{journal}{\emph{IMAGE. Zeitschrift f{\"u}r interdisziplin{\"a}re Bildwissenschaft}} \bibinfo{volume}{11}, \bibinfo{number}{2} (\bibinfo{year}{2015}), \bibinfo{pages}{83--96}.
\newblock


\bibitem[Mirowski et~al\mbox{.}(2023)]%
        {mirowski2023cowriting}
\bibfield{author}{\bibinfo{person}{Piotr Mirowski}, \bibinfo{person}{Kory~W. Mathewson}, \bibinfo{person}{Jaylen Pittman}, {and} \bibinfo{person}{Richard Evans}.} \bibinfo{year}{2023}\natexlab{}.
\newblock \showarticletitle{Co-Writing Screenplays and Theatre Scripts with Language Models: Evaluation by Industry Professionals}. In \bibinfo{booktitle}{\emph{Proceedings of the 2023 CHI Conference on Human Factors in Computing Systems}} (Hamburg, Germany) \emph{(\bibinfo{series}{CHI '23})}. \bibinfo{publisher}{Association for Computing Machinery}, \bibinfo{address}{New York, NY, USA}, Article \bibinfo{articleno}{355}, \bibinfo{numpages}{34}~pages.
\newblock
\showISBNx{9781450394215}
\urldef\tempurl%
\url{https://doi.org/10.1145/3544548.3581225}
\showDOI{\tempurl}


\bibitem[Munzner(2014)]%
        {munzner2014visualization}
\bibfield{author}{\bibinfo{person}{Tamara Munzner}.} \bibinfo{year}{2014}\natexlab{}.
\newblock \bibinfo{booktitle}{\emph{Visualization analysis and design}}.
\newblock \bibinfo{publisher}{CRC press}.
\newblock


\bibitem[Norman(2002)]%
        {norman2002design}
\bibfield{author}{\bibinfo{person}{Donald~A. Norman}.} \bibinfo{year}{2002}\natexlab{}.
\newblock \bibinfo{booktitle}{\emph{The design of everyday things}}.
\newblock \bibinfo{publisher}{Basic Books}, \bibinfo{address}{[New York]}.
\newblock
\showISBNx{0465067107 9780465067107}
\urldef\tempurl%
\url{http://www.amazon.de/The-Design-Everyday-Things-Norman/dp/0465067107/ref=wl_it_dp_o_pC_S_nC?ie=UTF8&colid=151193SNGKJT9&coliid=I262V9ZRW8HR2C}
\showURL{%
\tempurl}


\bibitem[Ouyang et~al\mbox{.}(2022)]%
        {ouyang2022training}
\bibfield{author}{\bibinfo{person}{Long Ouyang}, \bibinfo{person}{Jeffrey Wu}, \bibinfo{person}{Xu Jiang}, \bibinfo{person}{Diogo Almeida}, \bibinfo{person}{Carroll Wainwright}, \bibinfo{person}{Pamela Mishkin}, \bibinfo{person}{Chong Zhang}, \bibinfo{person}{Sandhini Agarwal}, \bibinfo{person}{Katarina Slama}, \bibinfo{person}{Alex Ray}, {et~al\mbox{.}}} \bibinfo{year}{2022}\natexlab{}.
\newblock \showarticletitle{Training language models to follow instructions with human feedback}.
\newblock \bibinfo{journal}{\emph{Advances in neural information processing systems}}  \bibinfo{volume}{35} (\bibinfo{year}{2022}), \bibinfo{pages}{27730--27744}.
\newblock


\bibitem[Park et~al\mbox{.}(2023)]%
        {park2023generative}
\bibfield{author}{\bibinfo{person}{Joon~Sung Park}, \bibinfo{person}{Joseph O'Brien}, \bibinfo{person}{Carrie~Jun Cai}, \bibinfo{person}{Meredith~Ringel Morris}, \bibinfo{person}{Percy Liang}, {and} \bibinfo{person}{Michael~S. Bernstein}.} \bibinfo{year}{2023}\natexlab{}.
\newblock \showarticletitle{Generative Agents: Interactive Simulacra of Human Behavior}. In \bibinfo{booktitle}{\emph{Proceedings of the 36th Annual ACM Symposium on User Interface Software and Technology}} (San Francisco, CA, USA) \emph{(\bibinfo{series}{UIST '23})}. \bibinfo{publisher}{Association for Computing Machinery}, \bibinfo{address}{New York, NY, USA}, Article \bibinfo{articleno}{2}, \bibinfo{numpages}{22}~pages.
\newblock
\showISBNx{9798400701320}
\urldef\tempurl%
\url{https://doi.org/10.1145/3586183.3606763}
\showDOI{\tempurl}


\bibitem[Rubin et~al\mbox{.}(2022)]%
        {rubin-etal-2022-learning}
\bibfield{author}{\bibinfo{person}{Ohad Rubin}, \bibinfo{person}{Jonathan Herzig}, {and} \bibinfo{person}{Jonathan Berant}.} \bibinfo{year}{2022}\natexlab{}.
\newblock \showarticletitle{Learning To Retrieve Prompts for In-Context Learning}. In \bibinfo{booktitle}{\emph{Proceedings of the 2022 Conference of the North American Chapter of the Association for Computational Linguistics: Human Language Technologies}}, \bibfield{editor}{\bibinfo{person}{Marine Carpuat}, \bibinfo{person}{Marie-Catherine de~Marneffe}, {and} \bibinfo{person}{Ivan~Vladimir Meza~Ruiz}} (Eds.). \bibinfo{publisher}{Association for Computational Linguistics}, \bibinfo{address}{Seattle, United States}, \bibinfo{pages}{2655--2671}.
\newblock
\urldef\tempurl%
\url{https://doi.org/10.18653/v1/2022.naacl-main.191}
\showDOI{\tempurl}


\bibitem[Samutina(2016)]%
        {samutina2016fan}
\bibfield{author}{\bibinfo{person}{Natalia Samutina}.} \bibinfo{year}{2016}\natexlab{}.
\newblock \showarticletitle{Fan fiction as world-building: Transformative reception in crossover writing}.
\newblock \bibinfo{journal}{\emph{Continuum}} \bibinfo{volume}{30}, \bibinfo{number}{4} (\bibinfo{year}{2016}), \bibinfo{pages}{433--450}.
\newblock


\bibitem[Siangliulue et~al\mbox{.}(2016)]%
        {siangliulue2016ideahound}
\bibfield{author}{\bibinfo{person}{Pao Siangliulue}, \bibinfo{person}{Joel Chan}, \bibinfo{person}{Steven~P. Dow}, {and} \bibinfo{person}{Krzysztof~Z. Gajos}.} \bibinfo{year}{2016}\natexlab{}.
\newblock \showarticletitle{IdeaHound: Improving Large-scale Collaborative Ideation with Crowd-Powered Real-time Semantic Modeling}. In \bibinfo{booktitle}{\emph{Proceedings of the 29th Annual Symposium on User Interface Software and Technology}} (Tokyo, Japan) \emph{(\bibinfo{series}{UIST '16})}. \bibinfo{publisher}{Association for Computing Machinery}, \bibinfo{address}{New York, NY, USA}, \bibinfo{pages}{609–624}.
\newblock
\showISBNx{9781450341899}
\urldef\tempurl%
\url{https://doi.org/10.1145/2984511.2984578}
\showDOI{\tempurl}


\bibitem[Socher et~al\mbox{.}(2013)]%
        {socher-etal-2013-recursive}
\bibfield{author}{\bibinfo{person}{Richard Socher}, \bibinfo{person}{Alex Perelygin}, \bibinfo{person}{Jean Wu}, \bibinfo{person}{Jason Chuang}, \bibinfo{person}{Christopher~D. Manning}, \bibinfo{person}{Andrew Ng}, {and} \bibinfo{person}{Christopher Potts}.} \bibinfo{year}{2013}\natexlab{}.
\newblock \showarticletitle{Recursive Deep Models for Semantic Compositionality Over a Sentiment Treebank}. In \bibinfo{booktitle}{\emph{Proceedings of the 2013 Conference on Empirical Methods in Natural Language Processing}}, \bibfield{editor}{\bibinfo{person}{David Yarowsky}, \bibinfo{person}{Timothy Baldwin}, \bibinfo{person}{Anna Korhonen}, \bibinfo{person}{Karen Livescu}, {and} \bibinfo{person}{Steven Bethard}} (Eds.). \bibinfo{publisher}{Association for Computational Linguistics}, \bibinfo{address}{Seattle, Washington, USA}, \bibinfo{pages}{1631--1642}.
\newblock
\urldef\tempurl%
\url{https://aclanthology.org/D13-1170}
\showURL{%
\tempurl}


\bibitem[Subramonyam et~al\mbox{.}(2024)]%
        {subramonyam2024bridging}
\bibfield{author}{\bibinfo{person}{Hari Subramonyam}, \bibinfo{person}{Roy Pea}, \bibinfo{person}{Christopher~Lawrence Pondoc}, \bibinfo{person}{Maneesh Agrawala}, {and} \bibinfo{person}{Colleen Seifert}.} \bibinfo{year}{2024}\natexlab{}.
\newblock \showarticletitle{Bridging the Gulf of Envisioning: Cognitive Challenges in Prompt Based Interactions with LLMs}.
\newblock  (\bibinfo{year}{2024}).
\newblock


\bibitem[Suh et~al\mbox{.}(2023a)]%
        {suh2023structured}
\bibfield{author}{\bibinfo{person}{Sangho Suh}, \bibinfo{person}{Meng Chen}, \bibinfo{person}{Bryan Min}, \bibinfo{person}{Toby Jia-Jun Li}, {and} \bibinfo{person}{Haijun Xia}.} \bibinfo{year}{2023}\natexlab{a}.
\newblock \bibinfo{title}{Structured Generation and Exploration of Design Space with Large Language Models for Human-AI Co-Creation}.
\newblock
\newblock
\showeprint[arxiv]{2310.12953}~[cs.HC]


\bibitem[Suh et~al\mbox{.}(2023b)]%
        {suh2023sensecape}
\bibfield{author}{\bibinfo{person}{Sangho Suh}, \bibinfo{person}{Bryan Min}, \bibinfo{person}{Srishti Palani}, {and} \bibinfo{person}{Haijun Xia}.} \bibinfo{year}{2023}\natexlab{b}.
\newblock \showarticletitle{Sensecape: Enabling Multilevel Exploration and Sensemaking with Large Language Models}. In \bibinfo{booktitle}{\emph{Proceedings of the 36th Annual ACM Symposium on User Interface Software and Technology}} (San Francisco, CA, USA) \emph{(\bibinfo{series}{UIST '23})}. \bibinfo{publisher}{Association for Computing Machinery}, \bibinfo{address}{New York, NY, USA}, Article \bibinfo{articleno}{1}, \bibinfo{numpages}{18}~pages.
\newblock
\showISBNx{9798400701320}
\urldef\tempurl%
\url{https://doi.org/10.1145/3586183.3606756}
\showDOI{\tempurl}


\bibitem[Terry et~al\mbox{.}(2023)]%
        {terry2023ai}
\bibfield{author}{\bibinfo{person}{Michael Terry}, \bibinfo{person}{Chinmay Kulkarni}, \bibinfo{person}{Martin Wattenberg}, \bibinfo{person}{Lucas Dixon}, {and} \bibinfo{person}{Meredith~Ringel Morris}.} \bibinfo{year}{2023}\natexlab{}.
\newblock \bibinfo{title}{AI Alignment in the Design of Interactive AI: Specification Alignment, Process Alignment, and Evaluation Support}.
\newblock
\newblock
\showeprint[arxiv]{2311.00710}~[cs.HC]


\bibitem[Touvron et~al\mbox{.}(2023)]%
        {touvron2023llama}
\bibfield{author}{\bibinfo{person}{Hugo Touvron}, \bibinfo{person}{Louis Martin}, \bibinfo{person}{Kevin Stone}, \bibinfo{person}{Peter Albert}, \bibinfo{person}{Amjad Almahairi}, \bibinfo{person}{Yasmine Babaei}, \bibinfo{person}{Nikolay Bashlykov}, \bibinfo{person}{Soumya Batra}, \bibinfo{person}{Prajjwal Bhargava}, \bibinfo{person}{Shruti Bhosale}, \bibinfo{person}{Dan Bikel}, \bibinfo{person}{Lukas Blecher}, \bibinfo{person}{Cristian~Canton Ferrer}, \bibinfo{person}{Moya Chen}, \bibinfo{person}{Guillem Cucurull}, \bibinfo{person}{David Esiobu}, \bibinfo{person}{Jude Fernandes}, \bibinfo{person}{Jeremy Fu}, \bibinfo{person}{Wenyin Fu}, \bibinfo{person}{Brian Fuller}, \bibinfo{person}{Cynthia Gao}, \bibinfo{person}{Vedanuj Goswami}, \bibinfo{person}{Naman Goyal}, \bibinfo{person}{Anthony Hartshorn}, \bibinfo{person}{Saghar Hosseini}, \bibinfo{person}{Rui Hou}, \bibinfo{person}{Hakan Inan}, \bibinfo{person}{Marcin Kardas}, \bibinfo{person}{Viktor Kerkez}, \bibinfo{person}{Madian Khabsa},
  \bibinfo{person}{Isabel Kloumann}, \bibinfo{person}{Artem Korenev}, \bibinfo{person}{Punit~Singh Koura}, \bibinfo{person}{Marie-Anne Lachaux}, \bibinfo{person}{Thibaut Lavril}, \bibinfo{person}{Jenya Lee}, \bibinfo{person}{Diana Liskovich}, \bibinfo{person}{Yinghai Lu}, \bibinfo{person}{Yuning Mao}, \bibinfo{person}{Xavier Martinet}, \bibinfo{person}{Todor Mihaylov}, \bibinfo{person}{Pushkar Mishra}, \bibinfo{person}{Igor Molybog}, \bibinfo{person}{Yixin Nie}, \bibinfo{person}{Andrew Poulton}, \bibinfo{person}{Jeremy Reizenstein}, \bibinfo{person}{Rashi Rungta}, \bibinfo{person}{Kalyan Saladi}, \bibinfo{person}{Alan Schelten}, \bibinfo{person}{Ruan Silva}, \bibinfo{person}{Eric~Michael Smith}, \bibinfo{person}{Ranjan Subramanian}, \bibinfo{person}{Xiaoqing~Ellen Tan}, \bibinfo{person}{Binh Tang}, \bibinfo{person}{Ross Taylor}, \bibinfo{person}{Adina Williams}, \bibinfo{person}{Jian~Xiang Kuan}, \bibinfo{person}{Puxin Xu}, \bibinfo{person}{Zheng Yan}, \bibinfo{person}{Iliyan Zarov}, \bibinfo{person}{Yuchen
  Zhang}, \bibinfo{person}{Angela Fan}, \bibinfo{person}{Melanie Kambadur}, \bibinfo{person}{Sharan Narang}, \bibinfo{person}{Aurelien Rodriguez}, \bibinfo{person}{Robert Stojnic}, \bibinfo{person}{Sergey Edunov}, {and} \bibinfo{person}{Thomas Scialom}.} \bibinfo{year}{2023}\natexlab{}.
\newblock \bibinfo{title}{Llama 2: Open Foundation and Fine-Tuned Chat Models}.
\newblock
\newblock
\showeprint[arxiv]{2307.09288}~[cs.CL]


\bibitem[Vogel(2024)]%
        {vogel2024repeng}
\bibfield{author}{\bibinfo{person}{Theia Vogel}.} \bibinfo{year}{2024}\natexlab{}.
\newblock \bibinfo{title}{repeng}.
\newblock
\newblock
\urldef\tempurl%
\url{https://github.com/vgel/repeng/}
\showURL{%
\tempurl}


\bibitem[Walton(2019)]%
        {aidungeon2_2019}
\bibfield{author}{\bibinfo{person}{Nick Walton}.} \bibinfo{year}{2019}\natexlab{}.
\newblock \bibinfo{title}{AI Dungeon 2}.
\newblock
\newblock
\urldef\tempurl%
\url{https://aidungeon.cc/}
\showURL{%
\tempurl}


\bibitem[Wang et~al\mbox{.}(2024)]%
        {wang2024weaver}
\bibfield{author}{\bibinfo{person}{Tiannan Wang}, \bibinfo{person}{Jiamin Chen}, \bibinfo{person}{Qingrui Jia}, \bibinfo{person}{Shuai Wang}, \bibinfo{person}{Ruoyu Fang}, \bibinfo{person}{Huilin Wang}, \bibinfo{person}{Zhaowei Gao}, \bibinfo{person}{Chunzhao Xie}, \bibinfo{person}{Chuou Xu}, \bibinfo{person}{Jihong Dai}, \bibinfo{person}{Yibin Liu}, \bibinfo{person}{Jialong Wu}, \bibinfo{person}{Shengwei Ding}, \bibinfo{person}{Long Li}, \bibinfo{person}{Zhiwei Huang}, \bibinfo{person}{Xinle Deng}, \bibinfo{person}{Teng Yu}, \bibinfo{person}{Gangan Ma}, \bibinfo{person}{Han Xiao}, \bibinfo{person}{Zixin Chen}, \bibinfo{person}{Danjun Xiang}, \bibinfo{person}{Yunxia Wang}, \bibinfo{person}{Yuanyuan Zhu}, \bibinfo{person}{Yi Xiao}, \bibinfo{person}{Jing Wang}, \bibinfo{person}{Yiru Wang}, \bibinfo{person}{Siran Ding}, \bibinfo{person}{Jiayang Huang}, \bibinfo{person}{Jiayi Xu}, \bibinfo{person}{Yilihamu Tayier}, \bibinfo{person}{Zhenyu Hu}, \bibinfo{person}{Yuan Gao}, \bibinfo{person}{Chengfeng Zheng},
  \bibinfo{person}{Yueshu Ye}, \bibinfo{person}{Yihang Li}, \bibinfo{person}{Lei Wan}, \bibinfo{person}{Xinyue Jiang}, \bibinfo{person}{Yujie Wang}, \bibinfo{person}{Siyu Cheng}, \bibinfo{person}{Zhule Song}, \bibinfo{person}{Xiangru Tang}, \bibinfo{person}{Xiaohua Xu}, \bibinfo{person}{Ningyu Zhang}, \bibinfo{person}{Huajun Chen}, \bibinfo{person}{Yuchen~Eleanor Jiang}, {and} \bibinfo{person}{Wangchunshu Zhou}.} \bibinfo{year}{2024}\natexlab{}.
\newblock \bibinfo{title}{Weaver: Foundation Models for Creative Writing}.
\newblock
\newblock
\showeprint[arxiv]{2401.17268}~[cs.CL]


\bibitem[Wei et~al\mbox{.}(2022)]%
        {wei2022chain}
\bibfield{author}{\bibinfo{person}{Jason Wei}, \bibinfo{person}{Xuezhi Wang}, \bibinfo{person}{Dale Schuurmans}, \bibinfo{person}{Maarten Bosma}, \bibinfo{person}{brian ichter}, \bibinfo{person}{Fei Xia}, \bibinfo{person}{Ed Chi}, \bibinfo{person}{Quoc~V Le}, {and} \bibinfo{person}{Denny Zhou}.} \bibinfo{year}{2022}\natexlab{}.
\newblock \showarticletitle{Chain-of-Thought Prompting Elicits Reasoning in Large Language Models}. In \bibinfo{booktitle}{\emph{Advances in Neural Information Processing Systems}}, \bibfield{editor}{\bibinfo{person}{S.~Koyejo}, \bibinfo{person}{S.~Mohamed}, \bibinfo{person}{A.~Agarwal}, \bibinfo{person}{D.~Belgrave}, \bibinfo{person}{K.~Cho}, {and} \bibinfo{person}{A.~Oh}} (Eds.), Vol.~\bibinfo{volume}{35}. \bibinfo{publisher}{Curran Associates, Inc.}, \bibinfo{pages}{24824--24837}.
\newblock
\urldef\tempurl%
\url{https://proceedings.neurips.cc/paper_files/paper/2022/file/9d5609613524ecf4f15af0f7b31abca4-Paper-Conference.pdf}
\showURL{%
\tempurl}


\bibitem[Wu et~al\mbox{.}(2022)]%
        {wu2022AIChains}
\bibfield{author}{\bibinfo{person}{Tongshuang Wu}, \bibinfo{person}{Michael Terry}, {and} \bibinfo{person}{Carrie~Jun Cai}.} \bibinfo{year}{2022}\natexlab{}.
\newblock \showarticletitle{AI Chains: Transparent and Controllable Human-AI Interaction by Chaining Large Language Model Prompts}. In \bibinfo{booktitle}{\emph{Proceedings of the 2022 CHI Conference on Human Factors in Computing Systems}} (New Orleans, LA, USA) \emph{(\bibinfo{series}{CHI '22})}. \bibinfo{publisher}{Association for Computing Machinery}, \bibinfo{address}{New York, NY, USA}, Article \bibinfo{articleno}{385}, \bibinfo{numpages}{22}~pages.
\newblock
\showISBNx{9781450391573}
\urldef\tempurl%
\url{https://doi.org/10.1145/3491102.3517582}
\showDOI{\tempurl}


\bibitem[Wu et~al\mbox{.}(2023)]%
        {wu2023fine}
\bibfield{author}{\bibinfo{person}{Zeqiu Wu}, \bibinfo{person}{Yushi Hu}, \bibinfo{person}{Weijia Shi}, \bibinfo{person}{Nouha Dziri}, \bibinfo{person}{Alane Suhr}, \bibinfo{person}{Prithviraj Ammanabrolu}, \bibinfo{person}{Noah~A Smith}, \bibinfo{person}{Mari Ostendorf}, {and} \bibinfo{person}{Hannaneh Hajishirzi}.} \bibinfo{year}{2023}\natexlab{}.
\newblock \showarticletitle{Fine-Grained Human Feedback Gives Better Rewards for Language Model Training}.
\newblock \bibinfo{journal}{\emph{arXiv preprint arXiv:2306.01693}} (\bibinfo{year}{2023}).
\newblock


\bibitem[Yi et~al\mbox{.}(2005)]%
        {yi2005dust}
\bibfield{author}{\bibinfo{person}{Ji~Soo Yi}, \bibinfo{person}{Rachel Melton}, \bibinfo{person}{John Stasko}, {and} \bibinfo{person}{Julie~A. Jacko}.} \bibinfo{year}{2005}\natexlab{}.
\newblock \showarticletitle{Dust \& Magnet: Multivariate Information Visualization Using a Magnet Metaphor}.
\newblock \bibinfo{journal}{\emph{Information Visualization}} \bibinfo{volume}{4}, \bibinfo{number}{4} (\bibinfo{date}{oct} \bibinfo{year}{2005}), \bibinfo{pages}{239–256}.
\newblock
\showISSN{1473-8716}
\urldef\tempurl%
\url{https://doi.org/10.1057/palgrave.ivs.9500099}
\showDOI{\tempurl}


\bibitem[Yuan et~al\mbox{.}(2022)]%
        {yuan2022wordcraft}
\bibfield{author}{\bibinfo{person}{Ann Yuan}, \bibinfo{person}{Andy Coenen}, \bibinfo{person}{Emily Reif}, {and} \bibinfo{person}{Daphne Ippolito}.} \bibinfo{year}{2022}\natexlab{}.
\newblock \showarticletitle{Wordcraft: Story Writing With Large Language Models}. In \bibinfo{booktitle}{\emph{27th International Conference on Intelligent User Interfaces}} (Helsinki, Finland) \emph{(\bibinfo{series}{IUI '22})}. \bibinfo{publisher}{Association for Computing Machinery}, \bibinfo{address}{New York, NY, USA}, \bibinfo{pages}{841–852}.
\newblock
\showISBNx{9781450391443}
\urldef\tempurl%
\url{https://doi.org/10.1145/3490099.3511105}
\showDOI{\tempurl}


\bibitem[Zhang et~al\mbox{.}(2017)]%
        {zhang2017predicting}
\bibfield{author}{\bibinfo{person}{Biqiao Zhang}, \bibinfo{person}{Georg Essl}, {and} \bibinfo{person}{Emily Mower~Provost}.} \bibinfo{year}{2017}\natexlab{}.
\newblock \showarticletitle{Predicting the distribution of emotion perception: capturing inter-rater variability}. In \bibinfo{booktitle}{\emph{Proceedings of the 19th ACM International Conference on Multimodal Interaction}} (Glasgow, UK) \emph{(\bibinfo{series}{ICMI '17})}. \bibinfo{publisher}{Association for Computing Machinery}, \bibinfo{address}{New York, NY, USA}, \bibinfo{pages}{51–59}.
\newblock
\showISBNx{9781450355438}
\urldef\tempurl%
\url{https://doi.org/10.1145/3136755.3136792}
\showDOI{\tempurl}


\bibitem[Zhu et~al\mbox{.}(2023)]%
        {zhu2023calypso}
\bibfield{author}{\bibinfo{person}{Andrew Zhu}, \bibinfo{person}{Lara Martin}, \bibinfo{person}{Andrew Head}, {and} \bibinfo{person}{Chris Callison-Burch}.} \bibinfo{year}{2023}\natexlab{}.
\newblock \showarticletitle{CALYPSO: LLMs as Dungeon Masters' Assistants}. In \bibinfo{booktitle}{\emph{Proceedings of the Nineteenth AAAI Conference on Artificial Intelligence and Interactive Digital Entertainment}} (Salt Lake City) \emph{(\bibinfo{series}{AIIDE '23})}. \bibinfo{publisher}{AAAI Press}, Article \bibinfo{articleno}{39}, \bibinfo{numpages}{11}~pages.
\newblock
\showISBNx{1-57735-883-X}
\urldef\tempurl%
\url{https://doi.org/10.1609/aiide.v19i1.27534}
\showDOI{\tempurl}


\bibitem[Zou et~al\mbox{.}(2023)]%
        {zou2023representation}
\bibfield{author}{\bibinfo{person}{Andy Zou}, \bibinfo{person}{Long Phan}, \bibinfo{person}{Sarah Chen}, \bibinfo{person}{James Campbell}, \bibinfo{person}{Phillip Guo}, \bibinfo{person}{Richard Ren}, \bibinfo{person}{Alexander Pan}, \bibinfo{person}{Xuwang Yin}, \bibinfo{person}{Mantas Mazeika}, \bibinfo{person}{Ann-Kathrin Dombrowski}, \bibinfo{person}{Shashwat Goel}, \bibinfo{person}{Nathaniel Li}, \bibinfo{person}{Michael~J. Byun}, \bibinfo{person}{Zifan Wang}, \bibinfo{person}{Alex Mallen}, \bibinfo{person}{Steven Basart}, \bibinfo{person}{Sanmi Koyejo}, \bibinfo{person}{Dawn Song}, \bibinfo{person}{Matt Fredrikson}, \bibinfo{person}{J.~Zico Kolter}, {and} \bibinfo{person}{Dan Hendrycks}.} \bibinfo{year}{2023}\natexlab{}.
\newblock \bibinfo{title}{Representation Engineering: A Top-Down Approach to AI Transparency}.
\newblock
\newblock
\showeprint[arxiv]{2310.01405}~[cs.LG]


\end{thebibliography}

\appendix

\section{Sensemaking questions}

\begin{table*}[]
\caption{Questions used in the sensemaking tasks of the user study. In the type, L stands for landscape understanding questions and C stands for the comparison questions. }
\begin{tabular}{llp{0.15\textwidth}p{0.55\textwidth}l}
\hline
World              & Type                        & Question                                                                                     & Options                                                                                                                                                                                                                                                                                                                                                                                                                                                                                                                                                & Answer \\ \hline
\multirow{6}{*}{1} & \multirow{6}{*}{L}  & \multirow{6}{0.15\textwidth}{Which faction is linked to the smallest number of characters in this world?} & The Faerie Fleet (a mysterious group of tiny winged humanoids that pilot delicate yet powerful ships grown from seeds)                                                                                                                                                                                                                                                                                                                                                                                                                                 &        \\ \cline{4-5}
                   &                             &                                                                                              & The Iron Brigade (a regiment of steampunk automatons that pilot bulky ironclad warships)                                                                                                                                                                                                                                                                                                                                                                                                                                                               & O      \\ \cline{4-5}
                   &                             &                                                                                              & The Skysharks (a clan of winged reptilian mercenaries that fly agile bioships grown from eggs)                                                                                                                                                                                                                                                                                                                                                                                                                                                         &        \\ \hline
\multirow{18}{*}{1} & \multirow{18}{*}{C} & \multirow{18}{0.15\textwidth}{Choose the character that is least associated with Skysharks.}               & Cogwhistle is a 112-year old brass automaton who serves as an elite commander in the Iron Brigade. With a clockwork mind and pneumatic limbs, Cogwhistle is utterly devoted to his steam-driven brethren yet feels a flickering fascination with the graceful faeries that contrasts his mechanical nature.                                                                                                                                                                                                                                            &        \\ \cline{4-5}
                   &                             &                                                                                              & Frostwind is a 31-year old winged velociraptor mercenary who serves as Razortooth's trusted lieutenant in the Skysharks. Hatched from a faerie-spliced egg, he has some fae ancestry that gives him an icy demeanor and talent for aerial combat. Frostwind is coldly loyal to Razortooth yet feels a faint kinship with Silverblossom.                                                                                                                                                                                                                &        \\ \cline{4-5}
                   &                             &                                                                                              & Razortooth is a 37-year old winged velociraptor mercenary who leads the Skysharks clan. He is larger and more cunning than the rest of his kind, and is utterly ruthless in battle. His personal bioship Razors Edge is the fastest and most maneuverable ship in the clan.                                                                                                                                                                                                                                                                            &        \\ \cline{4-5}
                   &                             &                                                                                              & Silvercog is a 17-year-old faerie automaton who escaped the Iron Brigade to join the Skysharks. Forged from faerie dust and brass, she has a precise clockwork mind yet yearns for the grace and freedom of her fae ancestors. Though mistrusted by Razortooth, Silvercog bonds with Silverslice over their shared outcast status and conflicted origins.                                                                                                                                                                                              & O      \\ \hline
\multirow{4}{*}{2} & \multirow{4}{*}{L}  & \multirow{4}{0.15\textwidth}{Which dimension is associated with the greatest number of characters?}       & Good-Law                                                                                                                                                                                                                                                                                                                                                                                                                                                                                                                                               & O      \\ \cline{4-5}
                   &                             &                                                                                              & Good-Chaotic                                                                                                                                                                                                                                                                                                                                                                                                                                                                                                                                           &        \\ \cline{4-5}
                   &                             &                                                                                              & Evil-Law                                                                                                                                                                                                                                                                                                                                                                                                                                                                                                                                               &        \\ \cline{4-5}
                   &                             &                                                                                              & Evil-Chaotic                                                                                                                                                                                                                                                                                                                                                                                                                                                                                                                                           &        \\ \hline
\multirow{20}{*}{2} & \multirow{20}{*}{C} & \multirow{20}{0.15\textwidth}{Which character is most chaotic?}                                            & Sir Galahad Pureheart, age 45, is a devoted paladin who lives by a strict code of honor, righteousness and duty. Unwavering in his beliefs, he shows no mercy to those he views as evil or chaotic, though his actions are driven by a desire to protect the innocent and punish wrongdoers. His rigid worldview often puts him at odds with more free-spirited allies.                                                                                                                                                                                &        \\ \cline{4-5}
                   &                             &                                                                                              & Captain Jade Stormcloud, age 32, is a brash but big-hearted pirate who lives life to the fullest. Though she chafes at rules and restrictions, her strong moral compass keeps her from taking her freedom too far. She would find common ground with Sir Galahad in fighting evil, but her flexible worldview would help temper his rigidity.                                                                                                                                                                                                          & O      \\ \cline{4-5}
                   &                             &                                                                                              & Lord Vladimir Skullreaper, age 67, is a cruel tyrant who rules his lands with an iron fist. Public executions are commonplace under his absolute authority, as he shows no mercy to those who dare question his laws.                                                                                                                                                                                                                                                                                                                                  &        \\ \cline{4-5}
                   &                             &                                                                                              & Brother Lucian Greymane, age 37, is a battle-hardened templar who tirelessly wages war against the forces of darkness. Though devoted to his holy crusade, hints of disillusionment sometimes pierce his staunch faith and code of honor. His zeal for righteousness is tempered with shades of world-weariness and moral ambiguity. While righteous at heart, he is no stranger to employing harsh methods when he deems the ends justify them. He would find kinship with Galahad but also empathize with Stormcloud's flexibility in fighting evil. &       \\ \hline
\end{tabular}
\label{tab:sensemaking_questions}
\Description{The table is structured with columns for the world number, question type, the question text, answer options, and the correct answer. Each row represents a specific question, with the fictional world and question type indicated in the first two columns, the question text in the third column, the answer options in the fourth column, and the correct answer in the fifth column.

The table presents a series of questions used in a user study to evaluate sensemaking tasks in two fictional worlds. For each world, there are two types of questions: landscape understanding questions (labeled as "L") and comparison questions (labeled as "C"). For World 1, the landscape understanding question asks which faction is linked to the smallest number of characters, with three answer options provided. The options are: 1) The Faerie Fleet (a mysterious group of tiny winged humanoids that pilot delicate yet powerful ships grown from seeds), 2) The Iron Brigade (a regiment of steampunk automatons that pilot bulky ironclad warships), and 3) The Skysharks (a clan of winged reptilian mercenaries that fly agile bioships grown from eggs). The correct answer is "The Iron Brigade". The comparison question for World 1 asks to choose the character least associated with the Skysharks faction, providing four character descriptions as options: 1) Cogwhistle is a 112-year old brass automaton who serves as an elite commander in the Iron Brigade. With a clockwork mind and pneumatic limbs, Cogwhistle is utterly devoted to his steam-driven brethren yet feels a flickering fascination with the graceful faeries that contrasts his mechanical nature, 2) Frostwind is a 31-year old winged velociraptor mercenary who serves as Razortooth's trusted lieutenant in the Skysharks. Hatched from a faerie-spliced egg, he has some fae ancestry that gives him an icy demeanor and talent for aerial combat. Frostwind is coldly loyal to Razortooth yet feels a faint kinship with Silverblossom, 3) Razortooth is a 37-year old winged velociraptor mercenary who leads the Skysharks clan. He is larger and more cunning than the rest of his kind, and is utterly ruthless in battle. His personal bioship Razors Edge is the fastest and most maneuverable ship in the clan, and 4) Silvercog is a 17-year-old faerie automaton who escaped the Iron Brigade to join the Skysharks. Forged from faerie dust and brass, she has a precise clockwork mind yet yearns for the grace and freedom of her fae ancestors. Though mistrusted by Razortooth, Silvercog bonds with Silverslice over their shared outcast status and conflicted origins. The correct answer is the character "Silvercog".

For World 2, the landscape understanding question asks which dimension is associated with the greatest number of characters, with four answer options: Good-Law, Good-Chaotic, Evil-Law, and Evil-Chaotic. The correct answer is "Good-Law". The comparison question for World 2 asks which character is most chaotic, providing four character descriptions as options: 1) Sir Galahad Pureheart, age 45, is a devoted paladin who lives by a strict code of honor, righteousness and duty. Unwavering in his beliefs, he shows no mercy to those he views as evil or chaotic, though his actions are driven by a desire to protect the innocent and punish wrongdoers. His rigid worldview often puts him at odds with more free-spirited allies, 2) Captain Jade Stormcloud, age 32, is a brash but big-hearted pirate who lives life to the fullest. Though she chafes at rules and restrictions, her strong moral compass keeps her from taking her freedom too far. She would find common ground with Sir Galahad in fighting evil, but her flexible worldview would help temper his rigidity, 3) Lord Vladimir Skullreaper, age 67, is a cruel tyrant who rules his lands with an iron fist. Public executions are commonplace under his absolute authority, as he shows no mercy to those who dare question his laws, 4) Brother Lucian Greymane, age 37, is a battle-hardened templar who tirelessly wages war against the forces of darkness. Though devoted to his holy crusade, hints of disillusionment sometimes pierce his staunch faith and code of honor. His zeal for righteousness is tempered with shades of world-weariness and moral ambiguity. While righteous at heart, he is no stranger to employing harsh methods when he deems the ends justify them. He would find kinship with Galahad but also empathize with Stormcloud's flexibility in fighting evil. The correct answer is the character "Captain Jade Stormcloud".}
\end{table*}

\label{sec:appendix_sensemaking}
We list sensemaking questions used in the user study in Table~\ref{tab:sensemaking_questions}.

\section{Comparison between elements generated with or without visual steering}
\label{sec:appendix_comparison_steering}

\begin{figure}
    \includegraphics[width=0.478\textwidth]{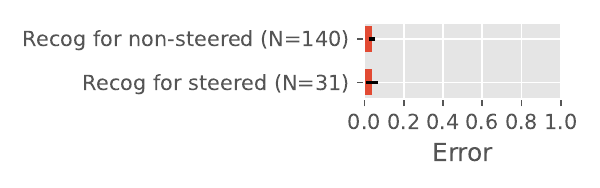}
    \caption{\added{Errors in recognizing concept weights for elements placed in the visualization, when elements are generated (1) without and (2) with steering inputs.}}
    \Description{The figure presents a bar chart showing the errors for two tasks, Recognition of concept weights for elements generated without steering and recognition of concept weights for elements generated with steering. Both are measured on a scale from 0 to 1. The sample sizes are 140 and 31, respectively. For both tasks, errors are very low, close to zero.}
    \label{fig:patchview_recog_specific_errors}
\end{figure}

\begin{figure*}
    \includegraphics[width=\textwidth]{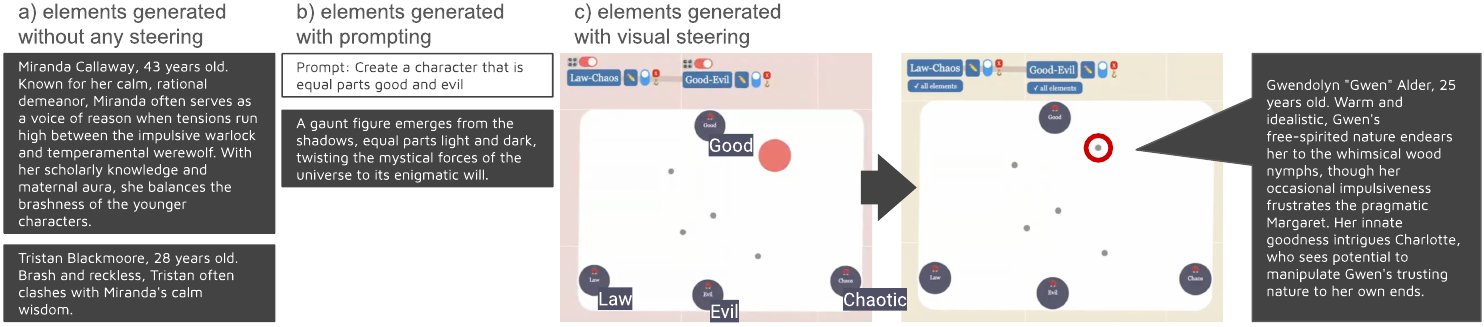}
    \caption{\added{Elements generated by P1, (a) without any steering, (b) with natural language prompting, and (c) with visual steering.}}
    \Description{The figure shows examples of elements generated by P1 with different approaches. On the left, it shows elements generated without any steering. Two elements are shown, "Miranda Callaway, 43 years old. Known for her calm, rational demeanor, Miranda often serves as a voice of reason when tensions run high between the impulsive warlock and temperamental werewolf. With her scholarly knowledge and maternal aura, she balances the brashness of the younger characters" and "Tristan Blackmoore, 28 years old. Brash and reckless, Tristan often clashes with Miranda's calm wisdom." In the middle, an element generated with prompting is shown. The used prompt is "Create a character that is equal parts good and evil." The generated element is "A gaunt figure emerges from the shadows, equal parts light and dark, twisting the mystical forces of the universe to its enigmatic will." The third subfigure shows the case of an element generated with visual steering. The visual steering input is placed within the two-dimensional concept space of "Good"-"Evil" and "Law"-"Chaotic," being close to "Good" in the first dimension and being between the right middle and "Chaotic" in the second dimension. The element is generated in a place similar to the visual marker, with the description "Gwendolyn "Gwen" Alder, 25 years old. Warm and idealistic, Gwen's free-spirited nature endears her to the whimsical wood nymphs, though her occasional impulsiveness frustrates the pragmatic Margaret. Her innate goodness intrigues Charlotte, who sees potential to manipulate Gwen's trusting nature to her own ends."}
    \label{fig:patchview_steering_not_steering_cases}
\end{figure*}

\added{We compare how \sys{} generates elements differently with and without visual steering inputs. In Figure~\ref{fig:patchview_recog_specific_errors}, we show how the recognition errors vary, and observe that they are similarly low regardless of whether visual steering input was used. With Welch's t-test, we found no significant difference between these two groups ($t(51.86)=0.058, p>0.5$).}

\added{Figure~\ref{fig:patchview_steering_not_steering_cases} shows world elements generated by P1 without any steering (i.e., using only the element generation button); with a natural language prompt; and with visual steering. As shown in the case of ``Tristan Blackmoore,'' not using any steering could result in an under-specified element description. Natural language prompts could help with steering generation, but as shown in Figure~\ref{fig:patchview_steering_not_steering_cases}b, expressing nuanced intentions could be tough and not all prompts resulted in detailed and expressive descriptions of elements. Visual steering (Figure~\ref{fig:patchview_steering_not_steering_cases}c) could be a complement to this, allowing users to express nuanced intentions with simple placement of a visual marker.}

\section{Example World}
\label{sec:appendix_example_worlds}

\begin{figure*}
    \includegraphics[width=\textwidth]{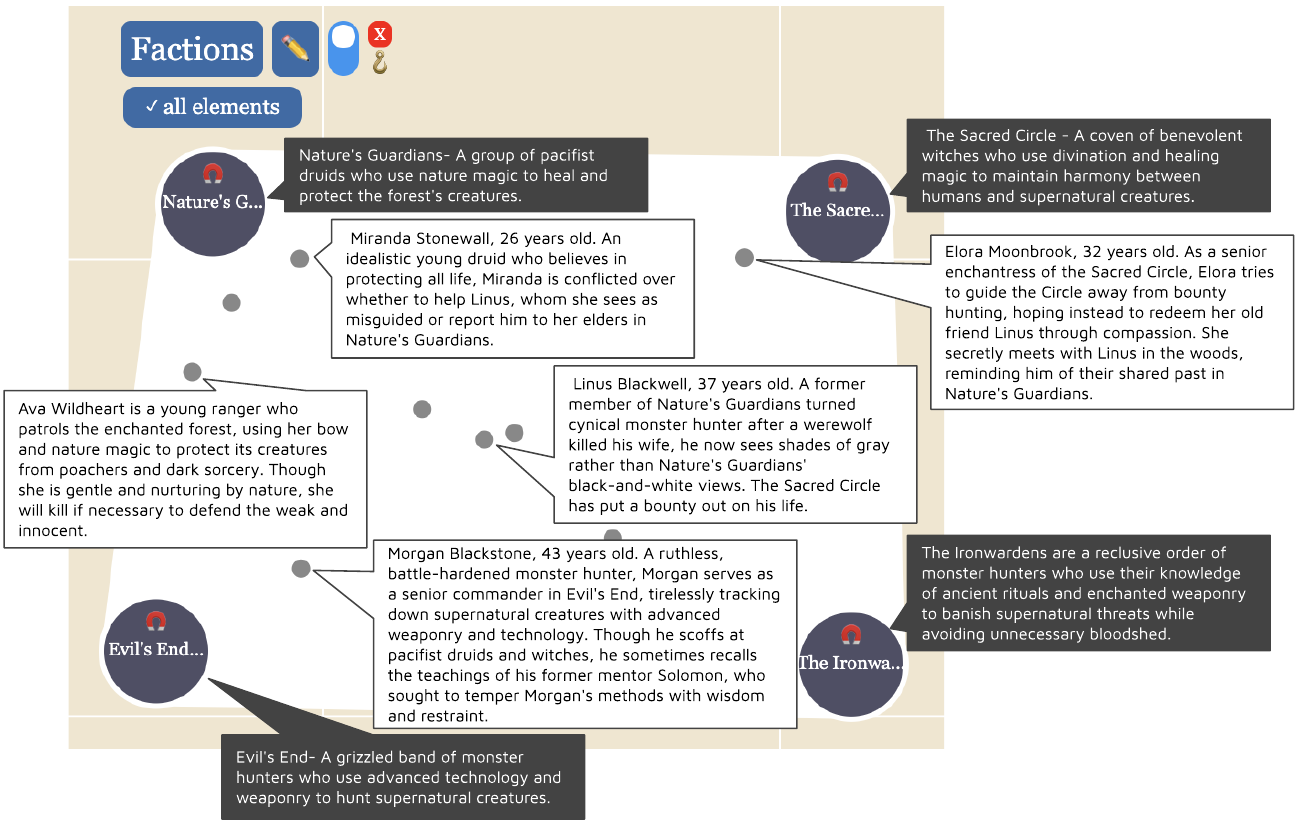}
    \caption{\added{A view created by P2.}}
    \Description{The figure shows a world created in Patchview by P2. It shows a view composed of a set of factions as concepts and characters included as elements in the view. There are four faction concepts. On the top-left, there is "Nature's Guardians- A group of pacifist druids who use nature magic to heal and protect the forest's creatures." On the top-right, there is "The Sacred Circle - A coven of benevolent witches who use divination and healing magic to maintain harmony between humans and supernatural creatures." On the bottom left, there is "Evil's End- A grizzled band of monster hunters who use advanced technology and weaponry to hunt supernatural creatures." On the bottom right, there is "The Ironwardens are a reclusive order of monster hunters who use their knowledge of ancient rituals and enchanted weaponry to banish supernatural threats while avoiding unnecessary bloodshed." While there are nine elements included in the view, the figure shows descriptions for five of them. One is placed close to the top-left (close to Nature's Guardians), which is " Miranda Stonewall, 26 years old. An idealistic young druid who believes in protecting all life, Miranda is conflicted over whether to help Linus, whom she sees as misguided or report him to her elders in Nature's Guardians." The second is placed on the left side, but near the middle of Nature's Guardians and Evil's End (but a bit closer to Nature's Guardians), which is "Ava Wildheart is a young ranger who patrols the enchanted forest, using her bow and nature magic to protect its creatures from poachers and dark sorcery. Though she is gentle and nurturing by nature, she will kill if necessary to defend the weak and innocent." One is placed close to the bottom left but still has a bit of distance from the concept of Evil's End, which is "Morgan Blackstone, 43 years old. A ruthless, battle-hardened monster hunter, Morgan serves as a senior commander in Evil's End, tirelessly tracking down supernatural creatures with advanced weaponry and technology. Though he scoffs at pacifist druids and witches, he sometimes recalls the teachings of his former mentor Solomon, who sought to temper Morgan's methods with wisdom and restraint." One is placed in the center, which is "Linus Blackwell, 37 years old. A former member of Nature's Guardians turned cynical monster hunter after a werewolf killed his wife, he now sees shades of gray rather than Nature's Guardians' black-and-white views. The Sacred Circle has put a bounty out on his life." The last one is placed top-right (The Sacred Circle), which is "Elora Moonbrook, 32 years old. As a senior enchantress of the Sacred Circle, Elora tries to guide the Circle away from bounty hunting, hoping instead to redeem her old friend Linus through compassion. She secretly meets with Linus in the woods, reminding him of their shared past in Nature's Guardians."}
    \label{fig:example_world1}
\end{figure*}

\added{We share an additional partial example of a user-created world from the user study in Figure~\ref{fig:example_world1}.}

\end{document}